\documentclass{amsart}
\usepackage{amsfonts}

\usepackage{amscd}
\usepackage{thmdefs}

\input{tcilatex}
\theoremstyle{definition}
\theoremstyle{remark}
\numberwithin{equation}{section}

\input tcilatex

\begin{document}
\title[Good spectral triples]{Good spectral triples, associated Lie groups of Campbell-Baker-Hausdorff
type and unimodularity}
\author{J. Marion*}
\author{K. Valavane}
\address{C. N. R. S.-Centre de Physique Th\'{e}orique de Marseille\\
Campus de Luminy, case 907\\
163, avenue de Luminy,\\
13288 Marseille Cedex 9, France }
\email{valavane@cpt.univ-mrs.fr}
\date{March, 1999}
\subjclass{22E65, 58B25, 46K10, 22D25}
\keywords{Good spectral triple, Lie group of Campbell-Baker-Hausdorff type, $\mathcal{K%
}$-topology, $\mathcal{K}$-trace, unimodularizable good spectral triple.\\
* deceased on September 1st 1998.}
\maketitle

\begin{abstract}
The notion of good spectral triple is initiated. We prove firstly that any
regular spectral triple may be embedded in a good spectral triple, so that,
in non-commutative geometry, we can restricts to deal only with good
spectral triples. Given a good spectral triple $\mathcal{K}=(\mathcal{A},%
\mathcal{H},D)$, we prove that $\mathcal{A}$ is naturally endowed with a
topology, called the $\mathcal{K}$-topology, making it into an unital
Fr\'{e}chet pr\'{e}-$C^{\ast }$-algebra, and that the group $Inv(\mathcal{A}%
) $ of its invertible elements has a canonical structure of Fr\'{e}chet Lie
group of Campbell-Baker-Hausdorff type open in its Lie algebra $\mathcal{A}$
; moreover, for any $n>0$ one has that $\mathcal{K}_{n}=(M_{n}(\mathcal{A}),%
\mathcal{H}\otimes \mathbb{C}^{n},D\otimes I_{n})$ is still a good spectral
triple. One deduces three important consequences. The first one concerns the 
$K$-theory of $M_{n}(\mathcal{A})$. The second is related to unitary and
sympletic groups : given any element $\Omega $ in $Inv(M_{n}(\mathcal{A}))$
such that $\Omega ^{\ast }=\Omega ^{-1}=\varepsilon \Omega ,\varepsilon \in
\{-1,1\}$, we prove that $U_{\Omega }(n,\mathcal{A})=\{a\in M_{n}(\mathcal{A}%
)/a^{\ast }.\Omega .a=a.\Omega .a^{\ast }=\Omega \}$ is a Fr\'{e}chet Lie
group with real Lie algebra $\mathcal{U}_{\Omega }(n,\mathcal{A})=\{x\in
M_{n}(\mathcal{A})/x^{\ast }.\Omega +\Omega .x=0\}$. Endly, we discuss about
unimodularizable good spectral triples $\mathcal{K}=(\mathcal{A},\mathcal{H}%
,D)$, that is to say those for which the algebra $\mathcal{A}$ may be
provided with a normalized trace $T$ which is continuous w.r.t. the $%
\mathcal{K}$-topology ; in such a case, with any integer $n>0$ we associate
a natural continuous trace $T_{n}$ on $M_{n}(\mathcal{A})$, and we prove
that $\mathcal{S}_{T}(n,\mathcal{A})=\{x\in M_{n}(\mathcal{A})/T_{n}(x)=0\}$
is the Lie algebra of a connected Fr\'{e}chet Lie subgroup $S_{T}(n,\mathcal{%
A})$ of $Inv(M_{n}(\mathcal{A}))$. Likewise, we prove that the Lie algebra $%
\mathcal{S}_{T}\mathcal{U}_{\Omega }(n,\mathcal{A})=\{x\in \mathcal{U}%
_{\Omega }(n,\mathcal{A})/T_{n}(x)=0\}$ is the Lie algebra of a connected
real Fr\'{e}chet subgroup $S_{T}U_{\Omega }(n,\mathcal{A})$ of $U_{\Omega
}(n,\mathcal{A})$.
\end{abstract}

\section{Introduction}

In infinite-dimensional analysis, it is well known that there is deep
breaking between the case of Banach spaces (over $\mathbb{R}$ or $\mathbb{C}$%
) and that of more general Hausdorff locally convex topological vector
spaces which are not normed spaces (see e.g. [2] or [12]). For example,
depending on the type of derivatives used (Fr\'{e}chet derivatives,
G\^{a}teaux derivatives,...), one obtains non equivalent
infinite-dimensional differential calculi, although each of them leads to a
Taylor expansion at a point ; unfortunately, for these various types of
differential calculi, the fundamental theorems of the Banach differential
calculus such that the ''Implicit function Theorem'' or the ''Frobenius
Theorem'' are no more valid. As a consequence, there are several theories of
''infinite-dimensional Lie groups'' depending on the class of groups and
underlying manifold structures we consider, (see e.g.[13]).

In the non-commutative geometry initiated by A Connes (see [5] and some
references therein), compact manifolds are replaced by unital $\ast $%
-algebras which are not assumed to be commutative. In this setting, one
obtains a quantized differential calculus as soon as one can incorporate
such an algebra $\mathcal{A}$ in a ''regular spectral triple'', that is to
say a triple $\mathcal{K}=(\mathcal{A},\mathcal{H},D)$ where $\mathcal{A}$
is represented by an unital $\ast $-subalgebra of the $\ast $-algebra $%
\mathcal{L}(\mathcal{H})$ of bounded operators on some separable Hilbert
space $\mathcal{H}$, and $D$ an unbounded self-adjoint operator on $\mathcal{%
H}$ with compact resolvent, such that for every element $a$ in $\mathcal{A}%
:[D,a]$ lies in $\mathcal{L}(\mathcal{H})$, and such that $\mathcal{A}$ is
contained in the domain $Dom(\delta ^{n})$ of the $n$-th power of the
derivation $\delta =[\sqrt{D.D},\cdot ]$ on $\mathcal{L}(\mathcal{H})$ for
every $n$ in $\mathbb{N}$.

A natural question, which has been already approached in the particular case
of gauge groups arising in non-commutative geometry in [8], [9] and [10],
can be then roughly formulated as follows : under what conditions the
quantized differential calculus associated with a regular spectral triple $%
\mathcal{K}=(\mathcal{A},\mathcal{H},D)$ leads to a tractable notion of
''infinite-dimensional Lie group'' for the group of invertible elements in $%
\mathcal{A}$ and some of its important subgroups? The present paper, which
brings an interesting and rather unexpected answer as soon as the spectral
triple fulfils some ''goodness'' property, is organized as follows.

-In section 1 we recall some basic definitions on smoothness and analyticity
in Fr\'{e}chet spaces and a group theoretical version of differentiability
yielding to the notion of generalized Lie group developed in [12], which, in
spite of the absence of an explicit underlying smooth manifold structure,
has the same main functional properties than a ''true'' Lie group.

-In section 2, with any regular spectral triple $\mathcal{K}=(\mathcal{A},%
\mathcal{H},D)$ we associate a countable family $(\left\| \cdot \right\|
_{n})_{n\in \mathbb{N}}$ of submultiplicative $*$-norms on $\mathcal{A}$
fulfilling a crucial property (Lemma 3.2). Denoting by $\mathcal{A}_{n}$ the
unital Banach $*$-algebra obtained as the completion of $\mathcal{A}$ with
respect to the norm $\left\| \cdot \right\| _{n}$, $n\in \mathbb{N}$, we
prove that the intersection $\mathcal{A}(\mathcal{K})$ of all the $\mathcal{A%
}_{n}$ is an unital Fr\'{e}chet $*$-algebra for the topology defined by the
family $(\left\| \cdot \right\| _{n})_{n\in \mathbb{N}}$, that we have
called the $\mathcal{K}$-topology of $\mathcal{A}(\mathcal{K})$.

-Section 3 is devoted to the notion of good spectral triple, which consists
of a regular spectral triple $\mathcal{K}=(\mathcal{A},\mathcal{H},D)$ for
which $\mathcal{A}(\mathcal{K})=\mathcal{A}$. It is proved that any regular
spectral triple may be canonically embedded in a good spectral triple, so
that, without loss of generality, given a regular spectral triple, we may
assume that it is good. Given a good spectral triple $\mathcal{K}=(\mathcal{A%
},\mathcal{H},D)$ with $\mathcal{A}$ endowed with the $\mathcal{K}$%
-topology, we prove that the product is analytic on $\mathcal{A}\times 
\mathcal{A}$ and that the involution is smooth on $\mathcal{A}$. We prove
also that the group $Inv(\mathcal{A})$ of invertible elements in $\mathcal{A}
$ has a natural structure of generalized Lie group with Lie algebra $%
\mathcal{A}$, and that the associated exponential mapping $Exp:\mathcal{A}%
\rightarrow Inv(\mathcal{A})$ is analytic.

-In section 4, we begin by recalling the notion of good algebra initiated by
R. Swan (see [14]) and extensively studied by J.-B. Bost (see [3]) : a
topological unital algebra $A$ is good if, for the induced topology, the
group of its invertible elements is a topological group open in $A$ ; the
unital Fr\'{e}chet good $\ast $-algebras have a holomorphic functional
calculus so that we can define on it the functors $K_{0}$ and $K_{1}$ of a
topological $K$-theory. Given a good spectral triple $\mathcal{K}=(\mathcal{A%
},\mathcal{H},D)$, using our crucial Lemma 2 and taking into account that $%
Inv(\mathcal{A})$ is a nice projective limit of analytic Banach Lie groups,
one obtains our main result (Theorem 5.3) : the group $Inv(\mathcal{A})$ has
a natural structure of analytic Fr\'{e}chet Lie group of
Campbell-Baker-Hausdorff type (in the sense of [6]-[13]) open in $\mathcal{A}
$ ; in particular, provided with the $\mathcal{K}$-topology, $\mathcal{A}$
is an unital Fr\'{e}chet good $\ast $-algebra.

We derive in Theorem 5.4 two important consequences : the embedding $i$ of $%
\mathcal{A}$ in the $C^{\ast }$-algebra completion $\mathcal{A}_{0}$ of $%
\mathcal{A}$ with respect to $\left\| .\right\| _{0}$ induces, at the level
of the $K$-theory, an isomorphism of groups $i_{\ast }:K_{j}(\mathcal{A}%
)\rightarrow K_{j}(\mathcal{A}_{0})$, $j=0,1$. Moreover, for $n>0$, $%
\mathcal{K}_{n}=(M_{n}(\mathcal{A}),\mathcal{H}\otimes \mathbb{C}%
^{n},D\otimes I_{n})$ is still a good spectral triple, so that the matrix
unital $\ast $-algebra $M_{n}(\mathcal{A})$ has a natural structure of
unital Fr\'{e}chet good $\ast $-algebra, improving then a result of R. Swan
([14], Corollary 1.2), and that $Inv(M_{n}(\mathcal{A}))$ is Fr\'{e}chet Lie
group of Campbell-Baker-Hausdorff type with Lie algebra $M_{n}(\mathcal{A})$.

-Section 5 uses extensively a result of J. Leslie and T. Robart (see [6],
[13]) which ensures the integrability of the closed Lie subalgebras of the
Lie algebra of a Lie group of Campbell-Baker-Hausdorff type. Given a good
spectral triple $\mathcal{K}=(\mathcal{A},\mathcal{H},D)$, we study first of
all subgroups of $Inv(M_{n}(\mathcal{A}))$ including pseudo-unitary and
symplectic groups ; more precisely, given any element $\Omega $ in $%
Inv(M_{n}(\mathcal{A}))$ such that $\Omega ^{-1}=\Omega ^{\ast }=\varepsilon
\Omega $ with $\varepsilon \in \{-1,1\}$, we prove that the group : 
\begin{equation*}
U_{\Omega }(n;\mathcal{A})=\{a\in M_{n}(\mathcal{A})/a^{\ast }.\Omega
.a=a.\Omega .a^{\ast }=\Omega \}
\end{equation*}
is a Fr\'{e}chet Lie of Campbell-Baker-Hausdorff type with Lie algebra 
\begin{equation*}
\mathcal{U}_{\Omega }(n;\mathcal{A})=\{x\in M_{n}(\mathcal{A})/x^{\ast
}.\Omega +\Omega .x=0\}.
\end{equation*}

Then, we discuss about unimodularizable good spectral triples $\mathcal{K}=(%
\mathcal{A},\mathcal{H},D)$, that is to say those for which $\mathcal{A}$
may be provided with a $\mathcal{K}$-trace. Clearly, we mean here any
normalized trace $T$ which is continuous with respect to the $\mathcal{K}$%
-topology of $\mathcal{A}$ ; in this case, with any integer $n>0$ we
associate a $\mathcal{K}_{n}$-trace $T_{n}$ on $M_{n}(\mathcal{A})$, we
prove that $\mathcal{S}_{T}(n;\mathcal{A})=\{x\in M_{n}(\mathcal{A}%
)/T_{n}(x)=0\}$ is the Lie algebra of a connected Fr\'{e}chet Lie subgroup $%
S_{T}(n;\mathcal{A})$ of the Fr\'{e}chet Lie group $Inv(M_{n}(\mathcal{A}))$%
, and that $\mathcal{S}_{T}\mathcal{U}_{\Omega }(n;\mathcal{A})=\{x\in 
\mathcal{U}_{\Omega }(n;\mathcal{A})/T_{n}(x)=0\}$ is the Lie algebra of a
connected real Fr\'{e}chet Lie subgroup $S_{T}U_{\Omega }(n;\mathcal{A})$ of 
$U_{\Omega }(n;\mathcal{A})$.

\section{Some preliminaries on analyticity and on generalized Lie groups}

As usual, $\mathbb{N}$ denotes the set on nonnegative integers, and we let : 
$\mathbb{N}^{*}=\mathbb{N}-\{0\}$.

\subsection{On smoothness and analyticity}

We recall here a brief outline of basic definitions related to smoothness
and analyticity in Fr\'{e}chet spaces (see e.g. [2] and [11]). Let $E$ and $%
F $ be two Fr\'{e}chet spaces over $\mathbb{C}$ or $\mathbb{R}$, let $E_{0}$
be a non empty open subset in $E$ and let $f$ be a continuous mapping from $%
E_{0}$ into $F$. Then :

(1) : $f$ is said to be of class $C^{1}$ on $E_{0}$ if for any $x$ in $E_{0}$
and any element $v$ in $E$ the G\^{a}teaux derivative $f^{\prime }(x;v)=%
\underset{t\rightarrow 0}{\lim }\frac{f(x+tv)-f(x)}{t}$ of $f$ at $x$ in the
direction $v$ exists, and if the mapping $f^{\prime }:(x,v)\in E_{0}\times
E\rightarrow f^{\prime }(x,v)\in F$ is continuous.

(2) : $f$ is said to be of class $C^{2}$ on $E_{0}$ if it is of class $C^{1}$
and if for each $v_{1}$ in $E$ the mapping $f_{v_{1}}^{\prime }:x\mapsto
f^{\prime }(x,v_{1})$ is of class $C^{1}$ on $E_{0}$ ; more generally, for $%
k\geq 1$, $f$ is of class $C^{k+1}$ on $E_{0}$, if it is of class $C^{k}$ on 
$E_{0}$ and if the mapping $f^{(k+1)}:$

\begin{equation*}
(x;v_{1},v_{2},...,v_{k+1})\longmapsto \underset{t\rightarrow 0}{\lim }\frac{%
f^{(k)}(x+tv_{k+1},v_{1},v_{2},...,v_{k})-f^{(k)}(x;v_{1},v_{2},...,v_{k})}{t%
}
\end{equation*}
is continuous on $E_{0}\times E^{k+1}$, and of class $C^{\infty }$ if for
any $k$ in $\mathbb{N}$, $f$ is of class $C^{k}$.

(3) : $f$ is said to be analytic on $E_{0}$ if $f$ is of class $C^{1}$ on $%
E_{0}$ and :

-in the case where $E$ and $F$ are complex vector spaces, if for any element 
$x$ in $E_{0}$ the mapping : $v\in E\mapsto f^{\prime }(x;v)\in F$ is a
continuous complex linear mapping from $E$ into $F$ ;

-in the case where $E$ and $F$ are real vector spaces, if there exists a non
empty open subset $W$ in $E$ such that $f$ may be extended into an analytic
mapping from $E_{0}\oplus iW$ into $F\oplus iF$.

Any analytic mapping $f$ from $E_{0}$ into $F$ is, of course, of class $%
C^{\infty }$ on $E_{0}$, and for any $k\geq 1$, any continuous $k$-linear
mapping from $E^{k}$ into $F$ is analytic.

\subsection{On the notion of generalized Lie group}

\begin{definition}
By topological group of exponential type, it is meant a triple $(G,\mathcal{G%
},E)$ in which $G$ is a metric group, $\mathcal{G}$ is a complete locally
convex topological vector space and $E$ is a continuous mapping from $%
\mathcal{G}$ into $G$ such that :

(1) : $\forall v\in \mathcal{G}$, the mapping $\psi _{v}:t\in \mathbb{R}%
\mapsto \psi _{v}(t)=E(tv)\in G$ is a 1-parameter subgroup of $G$ ;

(2) : given any pair $(v,w)$ of elements in $\mathcal{G}$, one has $%
v=w\Leftrightarrow \psi _{v}=\psi _{w}$ ;

(3) : a sequence $(v_{n})$ of elements in $\mathcal{G}$ converges to an
element $v$ in $\mathcal{G}$ if and only if the corresponding sequence of
elements $(\psi _{v_{n}})$ of 1-parameter subgroups converges uniformly on
each compact interval to the 1-parameter subgroup $\psi _{v}$ ;

(4) : there exists a continuous mapping $\alpha :G\times \mathcal{G}%
\rightarrow \mathcal{G}$ fulfilling : 
\begin{equation*}
a.E(tv).a^{-1}=E(t\alpha (a,v))
\end{equation*}
for any element $a$ in $G$ and any element $v$ in $\mathcal{G}$.
\end{definition}

b) The notion of topological group of exponential type (see [12], chap 1.2)
gives a coherent setting for defining a group theoretical version of
differentiability.

Given a topological group of exponential type $(G,\mathcal{G},E)$ and a
continuous mapping $U$ from $\mathbb{R}$ into $G$, for any integer $n\geq 0$
and any real number $t$, let $F_{t,n}:\mathbb{R}\rightarrow G$ be the
mapping defined for all $s$ in $\mathbb{R}$ by : $F_{t,n}(s)=(U(t+\frac{s}{n}%
).U(t)^{-1})^{n}$ ; then :

\begin{definition}
(1) : $U$ is said to be differentiable at $t$ if there exist an element $%
X(t) $ in $\mathcal{G}$, called the derivation of $U$ at $t$ and denoted
usually by $\partial _{t}^{\log }U$, such that on each compact interval the
sequence $(F_{t,n})_{n}$ converges uniformly to the 1-parameter subgroup $%
E(sX(t))$.

(2) : $U$ will be said of class $C^{1}$, or a $C^{1}$-curve in $G$, if $U$
is differentiable at any $t$ and if the mapping : $t\in \mathbb{R}%
\rightarrow \partial _{t}^{\log }U\in \mathcal{G}$ is continuous.. We denote
by $\mathbb{I}$\textbf{\ }the unit in\textbf{\ }$G$, by $C^{1}(G,\mathcal{G}%
) $ the space of its $C^{1}$-curves and by $C_{\mathbb{I}}^{1}(G,\mathcal{G}%
) $ the subset : $C_{\mathbb{I}}^{1}(G,\mathcal{G})=\{U\in C^{1}(G,\mathcal{G%
})/U(0)=\mathbb{I}\}$.
\end{definition}

c) According to [12], chap 1.3, we are now able to say what is a generalized
Lie group. Let $I$ be a closed interval $[a,b]$, let $S(I;\mathcal{G})$
associated with some subdivision $\Delta :a=t_{0}<t_{1}<..<t_{m}=b$ of $I$,
and let us consider the continuous mapping :

\begin{equation*}
P(X):t\in I\longmapsto \prod_{a}^{t}E(X(s))ds\in G
\end{equation*}
called a product integral, and defined for any element $t$ in $%
[t_{k},t_{k+1}[$, $0\leq k\leq m-1$ by :

\begin{equation*}
P(X)(t)=%
\prod_{a}^{t}E(X(s))ds=E((t-t_{k})X(t_{k})).E((t_{k}-t_{k-1})X(t_{k-1})).%
\cdot \cdot \cdot .E((t_{1}-t_{0})X(t_{0}))
\end{equation*}

Such a mapping is a piecewise $C^{1}$-curve in $G$, and fulfils : 
\begin{equation*}
\partial _{t}^{\log }(P(X))=X(t_{k})\quad \forall t\in [t_{k},t_{k+1}[.
\end{equation*}

\begin{definition}
Let $(G,\mathcal{G},E)$ be a topological group of exponential type. $G$ is
called a generalized Lie group with Lie algebra $\mathcal{G}$ and
exponential mapping $E$ if :

(L1) : $C^{1}(G,\mathcal{G})$ is a group under the pointwise product, and
for any pair $(f,g)$ of elements in $C_{\mathbb{I}}^{1}(G,\mathcal{G})$ one
has : $\partial _{t}^{\log }(f.g)=\partial _{t}^{\log }(f)(0)+\partial
_{t}^{\log }(g)(0)$.

(L2) : For any continuous mapping $X:I\rightarrow \mathcal{G}$ and any
sequence $(X_{n})$ of elements in $S(I;\mathcal{G})$ converging uniformly to 
$X$, the sequence $(P(X_{n}))$ of product integrals converges uniformly, on
each compact interval, to a $C^{1}$-curve $g$ in $G$ such that $\partial
_{t}^{\log }(g)=X(t),t\in I$.

(L3) : For any $f$ in $C_{\mathbb{I}}^{1}(G,\mathcal{G})$ and any $v$ in $%
\mathcal{G}$, the mapping $t\mapsto \alpha (f(t),v)$ is differentiable on $%
\mathbb{R}$ and $\frac{d}{dt}(\alpha (f(t),v))_{t=0}=\frac{d}{dt}(\alpha
(E(t\partial _{t}^{\log }f(0),v))$ for any $f$ in $C_{\mathbb{I}}^{1}(G,%
\mathcal{G})$.

(L4) : The mapping $[.,.]:(u,v)\in \mathcal{G}\times \mathcal{G}\mapsto
[u,v]=\frac{d}{dt}(\alpha (E(tu),v))_{t=0}$ is continuous.
\end{definition}

\section{Regular spectral triples}

\subsection{On the notion of regular spectral triple}

\begin{definition}
A regular spectral triple is a triplet $\mathcal{K}=(\mathcal{A},\mathcal{H}%
,D)$ in which :

(1) : $\mathcal{A}$ is an unital involutive subalgebra, with unit denoted by 
$\mathbb{I}$ and involution by $*$, of the $*$-algebra $\mathcal{L}(\mathcal{%
H}) $ of bounded linear operators on some separable Hilbert space $\mathcal{H%
}$ ;

(2) : $D$ is an unbounded self-adjoint operator on $\mathcal{H}$ with
compact resolvent, and such that for any element $a$ in $D$ the operator $%
[D,a]=D.a-a.D$ lies in $\mathcal{L}(\mathcal{H})$ ;

(3) : $\mathcal{A}$ is contained in the domain $Dom(\delta ^{n})$ of the $n$%
-th power of the derivation $\delta =[\left| D\right| ,.]$ on $\mathcal{L}(%
\mathcal{H})$ for every $n$ in $\mathbb{N}$, where $\left| D\right| =\sqrt{%
D.D}$.
\end{definition}

According to [5], conditions (1) and (2) are the properties required in
order that $\mathcal{K}$ is a $K$-cycle over $\mathcal{A}$ ; the smoothness
condition (3) ensures that, for any a in $\mathcal{A}$ and any $n\geq 0$,
the $n$-th derivative of $a$ with respect to the derivation $\partial =[D,.]$
is well defined.\\

Given a regular unital spectral triple $\mathcal{K}=(\mathcal{A},\mathcal{H}%
,D)$ we shall denote by $\left\| .\right\| $ the $C^{*}$-norm operator on $%
\mathcal{L}(\mathcal{H})$, and for each integer $n\geq 0$, by $T^{(n)}$ the
family : $T^{(n)}=(T_{k})_{0\leq k\leq n}$, where $T_{k}$ is the seminorm
defined on $\mathcal{A}$ by :

\begin{equation*}
\left\{ 
\begin{array}{l}
T_{0}(a)=\left\| a\right\| \\ 
T_{k}(a)=\frac{1}{k!}\left\| \partial ^{k}(a)\right\| ,1\leq k\leq n
\end{array}
\right\} .
\end{equation*}

Endly, we shall set :

\begin{equation*}
\left\| .\right\| _{n}=\underset{0\leq k\leq n}{\sum }T_{k}.
\end{equation*}

\begin{lemma}
For any natural integer $n$ the following properties hold :

(i) : For any pair $(a,b)$ of elements in $\mathcal{A}$ one has : $\left\|
a.b\right\| _{n}\leq \left\| a\right\| _{n}\left\| b\right\| _{n}$.

(ii) : For any element $a$ in $\mathcal{A}$ one has : $\left\| a^{*}\right\|
_{n}=\left\| a\right\| _{n}$.

(iii) : For any element $a$ in $\mathcal{A}$ one has : $\left\| a\right\|
_{n+1}=\left\| a\right\| _{n}+\frac{1}{(n+1)!}\left\| \partial
^{(n+1)}(a)\right\| $.
\end{lemma}

\proof%
%
(i) : One easily checks that for each integer $n$, $T^{(n)}$ is a
differential seminorm on $\mathcal{A}$ ; moreover, from Proposition 3.3 of
[1] its associated total norm $\left\| .\right\| _{n}$ is submultiplicative.

(ii) : A trivial computation gives, for any positive integer $k:$%
\begin{equation*}
\delta ^{k}(a^{*})=(-1)^{k}(\delta ^{k}(a))^{*}
\end{equation*}
and then, since $\left\| .\right\| $ is $C^{*}$-norm :

\begin{equation*}
T_{k}(a^{*})=T_{k}(a)\text{, and }\left\| a^{*}\right\| _{n}=\left\|
a\right\| _{n}.
\end{equation*}

(iii) is an obvious consequence of the definition of $\left\| .\right\| _{n}$%
\endproof%
%

\begin{lemma}
Let $a$ be any element in $\mathcal{A}$. For any element $n$ in $\mathbb{N}%
-\{0\}$ there exists a positive constant $\eta _{a,n}$, such that $\left\|
a.x\right\| _{n}\leq \left\| a\right\| _{0}\left\| x\right\| _{n}+\eta
_{a,n}\left\| x\right\| _{n-1}$, $\forall x\in \mathcal{A}$.
\end{lemma}

\proof%
%
Let us prove firstly the statement for $n=1$. For any pair $(a,x)$ of
elements in $\mathcal{A}$, by \S\ 2.1(b) one has :

\begin{equation*}
\left\| a.x\right\| _{1}=\left\| a.x\right\| _{0}+\left\| [D,a.x]\right\|
\leq \left\| a\right\| _{0}\left\| x\right\| _{0}+\left\| [D,a]\right\|
\left\| x\right\| +\left\| a\right\| \left\| [D,x]\right\| ,
\end{equation*}
so that :

\begin{equation*}
\left\| a.x\right\| _{1}\leq \left\| a\right\| _{0}\left( \left\| x\right\|
_{0}+\left\| [D,x]\right\| \right) +\left\| [D,a]\right\| \left\| x\right\|
_{0},
\end{equation*}
and then :

\begin{equation*}
\left\| a.x\right\| _{1}\leq \left\| a\right\| _{0}\left\| x\right\|
_{1}+\left\| [D,a]\right\| \left\| x\right\| _{0}.
\end{equation*}

The assertion is then proved by taking $\eta _{a,1}=\left\| [D,a]\right\| $.

Now, let $k$ be an integer \TEXTsymbol{>}1, and let us assume that there
exists a positive constant $\eta _{a,k}$ such that for any pair $(a,x)$ of
elements in $\mathcal{A}:$

\begin{equation*}
\left\| a.x\right\| _{k}\leq \left\| a\right\| _{0}\left\| x\right\|
_{k}+\eta _{a,k}\left\| x\right\| _{k-1}.
\end{equation*}

From the equality :

\begin{equation*}
\left\| a.x\right\| _{k+1}=\left\| a.x\right\| _{k}+\frac{1}{(k+1)!}\left\|
\partial ^{k+1}(a.x)\right\| ,
\end{equation*}
the Leibniz rule leads to :

\begin{equation*}
\left\| a.x\right\| _{k+1}\leq \left\| a.x\right\| _{k}+\frac{1}{(k+1)!}%
\left\| \sum_{0\leq j\leq k+1}\left( 
\begin{array}{c}
k+1 \\ 
j
\end{array}
\right) \partial ^{j}(a).\partial ^{k+1-j}(x)\right\| ,
\end{equation*}
and then :

\begin{equation*}
\left\| a.x\right\| _{k+1}\leq \left\| a.x\right\| _{k}+\frac{1}{(k+1)!}%
\left( \left\| a\right\| _{k}\left\| \partial ^{k+1}(x)\right\| +\left\|
\sum_{1\leq j\leq k+1}\left( 
\begin{array}{c}
k+1 \\ 
j
\end{array}
\right) \partial ^{j}(a).\partial ^{k+1-j}(x)\right\| \right)
\end{equation*}
so that, by inductive assumption and the equality :

\begin{equation*}
\left\| x\right\| _{k+1}=\left\| x\right\| _{k}+\frac{1}{(k+1)!}\left\|
\partial ^{k+1-j}(x)\right\| ,
\end{equation*}
one obtains :

\begin{equation*}
\left\| a.x\right\| _{k+1}\leq A_{k}+B_{k}+C_{k},
\end{equation*}
with :

\begin{equation*}
\left\{ 
\begin{array}{l}
A_{k}=\left\| a\right\| _{0}\left\| x\right\| _{k+1} \\ 
B_{k}=\eta _{a,k}\left\| x\right\| _{k-1} \\ 
C_{k}=\frac{1}{(k+1)!}\underset{1\leq j\leq k+1}{\sum }\left( 
\begin{array}{c}
k+1 \\ 
j
\end{array}
\right) \left\| \partial ^{j}(a)\right\| \left\| \partial
^{k+1-j}(x)\right\| .
\end{array}
\right.
\end{equation*}

Let $\alpha (a,k)$ be the positive number defined by :

\begin{equation*}
\alpha (a,k)=\underset{1\leq j\leq k+1}{\max }\left\| \left( 
\begin{array}{c}
k+1 \\ 
j
\end{array}
\right) \partial ^{j}(a)\right\| ;
\end{equation*}

one deduces that :

\begin{equation*}
C_{k}\leq \leq \alpha (a,k)\left( \sum_{0\leq j\leq k+1}\frac{1}{(k+1)!}%
\left\| \partial ^{k+1-j}(x)\right\| \right) \leq \alpha (a,k)\left(
\sum_{0\leq j\leq k+1}\frac{1}{(k+1-j)!}\left\| \partial ^{k+1-j}(x)\right\|
\right) .
\end{equation*}

For any integer $j$ such that $1\leq j\leq k+1$ one has :

\begin{equation*}
\frac{1}{(k+1-j)!}\left\| \partial ^{k+1-j}(x)\right\| =\left\| x\right\|
_{k+1-j}-\left\| x\right\| _{k-j},
\end{equation*}
from which one deduces that $C_{k}\leq \alpha (a,k)\left\| x\right\| _{k}$,
and then, by (3) :

\begin{equation*}
\left\| a.x\right\| _{k+1}\leq \left\| a\right\| _{0}\left\| x\right\|
_{k+1}+\eta _{a,k}\left\| x\right\| _{k-1}+\alpha (a,k)\left\| x\right\|
_{k}.
\end{equation*}

Endly, using the fact that $\left\| x\right\| _{k-1}\leq \left\| x\right\|
_{k}$, one obtains :

\begin{equation*}
\left\| a.x\right\| _{k+1}\leq \left\| a\right\| _{0}\left\| x\right\|
_{k+1}+(\eta _{a,k}+\alpha (a,k))\left\| x\right\| _{k},
\end{equation*}
which achieves the proof%
\endproof%
%

\subsection{Unital Fr\'{e}chet $*$-algebra associated with a regular
spectral triple}

a) Let us recall what is called an $ILB$-chain (see e.g. [11], Chap I) : a
family $\{E,E_{n};n\in \mathbb{N}\}$ is called an $ILB$-chain if the
following conditions are satisfied :

-$\forall n\in \mathbb{N}:E_{n}$ is a Banach space (over $\mathbb{C}$ or $%
\mathbb{R}$), $E_{n+1}$ is embedded continuously in $E_{n}$ and the
embedding $i_{n}:E_{n+1}\hookrightarrow E_{n}$ is a continuous morphism of
Banach spaces with dense range ;

-$E=\underset{n\geq 0}{\bigcap }E_{n}$ is a Fr\'{e}chet space and the
topology of $E$ is the weakest topology such that the natural embedding $%
i_{n}:E_{n+1}\hookrightarrow E_{n}$ is a continuous for all $n$ in $\mathbb{N%
}$.\\

Given a regular spectral triple $\mathcal{K}=(\mathcal{A},\mathcal{H},D)$,
it follows from \S\ 2.1 that $\left\| .\right\| _{0}$ is a $C^{*}$-norm and
that $\left( \left\| .\right\| _{n}\right) _{n\in \mathbb{N}}$ is a
countable increasing family of submultiplicative $*$-norms on $\mathcal{A}$.
The completion $\mathcal{A}_{n}$ of $\mathcal{A}$ with respect to $\left\|
.\right\| _{n}$ is then an unital Banach $*$-algebra, $\mathcal{A}_{0}$
being a $C^{*}$-algebra. By Lemma 3.1, for any element $n$ in $\mathbb{N}$,
one has that $\mathcal{A}_{n+1}$ is a dense unital $*$-subalgebra of $%
\mathcal{A}_{n}$ and that the embedding $i_{n}:\mathcal{A}%
_{n+1}\hookrightarrow \mathcal{A}_{n}$ is a continuous morphism of Banach $*$%
-algebras with dense range.

One easily deduces that :

\begin{proposition}
Let $\mathcal{K}=(\mathcal{A},\mathcal{H},D)$ be a regular spectral triple.

The family $\{\mathcal{A}_{n},\left\| .\right\| _{n},i_{n},n\in \mathbb{N}\}$
is a projective system of unital involutive Banach algebras, and its
projective limit :

\begin{equation*}
\mathcal{A}(\mathcal{K})=\underset{\longleftarrow }{Lim}\mathcal{A}_{n}=%
\underset{n\geq 0}{\bigcap }\mathcal{A}_{n}
\end{equation*}
is an unital Fr\'{e}chet $*$-algebra the topology of which, called the $%
\mathcal{K}$-topology, is given by the family of $*$-norms $\left( \left\|
.\right\| _{n}\right) _{n\in \mathbb{N}}$. In particular, $\{\mathcal{A}(%
\mathcal{K}),\mathcal{A}_{n},n\in \mathbb{N}\}$ is an $ILB$-chain.
\end{proposition}

\section{Good spectral triple}

\subsection{On the notion of good spectral triple}

\begin{definition}
By good spectral triple, it will be meant any regular spectral triple $%
\mathcal{K}=(\mathcal{A},\mathcal{H},D)$ such that $\mathcal{A}=\mathcal{A}(%
\mathcal{K})$.
\end{definition}

\begin{proposition}
Let $\mathcal{K}=(\mathcal{A},\mathcal{H},D)$ be a regular spectral triple.
Then : 
\begin{equation*}
\mathcal{K}^{\#}=(\mathcal{A}(\mathcal{K}),\mathcal{H},D)
\end{equation*}
is good spectral triple.
\end{proposition}

\proof%
%
Given any spectral triple $\mathcal{K}=(\mathcal{A},\mathcal{H},D)$, and
taking into account that on $\mathcal{A}$ the $\mathcal{K}$-topology is
stronger that the topology induced by the $C^{*}$-norm $\left\| .\right\| $,
the inclusion $a\in \mathcal{A}\mapsto a\in \mathcal{L}(\mathcal{H})$ is
continuous on $\mathcal{A}$ provided with the $\mathcal{K}$-topology, and
then, extends into the continuous inclusion from $\mathcal{A}(\mathcal{K})$
into $\mathcal{L}(\mathcal{H})$. One easily deduces that $\mathcal{K}^{\#}=(%
\mathcal{A}(\mathcal{K}),\mathcal{H},D)$, like $\mathcal{K}$, is a regular
spectral triple. Now, observing that for any integer $n\geq 0$ one has
necessarily $\mathcal{A}(\mathcal{K})_{n}=\mathcal{A}_{n}$, one deduces that
:

\begin{equation*}
\mathcal{A}(\mathcal{K})(\mathcal{K}^{\#})=\underset{\longleftarrow }{Lim}%
\mathcal{A}(\mathcal{K})_{n}=\underset{n\geq 0}{\bigcap }\mathcal{A}(%
\mathcal{K})_{n}=\underset{n\geq 0}{\bigcap }\mathcal{A}_{n}=\mathcal{A}(%
\mathcal{K}),
\end{equation*}
which proves that $\mathcal{K}^{\#}=(\mathcal{A}(\mathcal{K}),\mathcal{H},D)$
is good spectral triple%
\endproof%
%

\textbf{Remark 4.1. }By Proposition 4.1, with any regular spectral triple $%
\mathcal{K}=(\mathcal{A},\mathcal{H},D)$ is associated the good spectral
triple $\mathcal{K}^{\#}=(\mathcal{A}(\mathcal{K}),\mathcal{H},D)$. $%
\mathcal{A}$ being a continuously and densely embedded unital $*$-subalgebra
of $\mathcal{A}(\mathcal{K})$ for the $\mathcal{K}$-topology which is
stronger than the weak topology, one has :

\begin{equation*}
\mathcal{A}\subseteq \mathcal{A}(\mathcal{K})\subseteq \mathcal{A}^{\prime
\prime },
\end{equation*}
where $\mathcal{A}^{\prime \prime }$ denotes the bicommutant of $\mathcal{A}$%
. It follows that, according to the ideas of A. Connes ([5]), replacing $%
\mathcal{A}$ by $\mathcal{A}(\mathcal{K})$ if necessary, there is no loss of
generality to consider only good spectral triples. As a Corollary of Lemma
3.1 and Proposition 3.3 one obtains :

\begin{proposition}
Let $\mathcal{K}=(\mathcal{A},\mathcal{H},D)$ be a good spectral triple.
Provided with its $\mathcal{K}$-topology, $\mathcal{A}$ is an unital
Fr\'{e}chet $*$-algebra, the involution $*$ is continuous on $\mathcal{A}$,
and the multiplication is continuous on $\mathcal{A}\times \mathcal{A}$.
\end{proposition}

\subsection{Analyticity of the product and smoothness of the involution}

\begin{proposition}
Let $\mathcal{K}=(\mathcal{A},\mathcal{H},D)$ be a good spectral triple. $%
\mathcal{A}$ being provided with the $\mathcal{K}$-topology, one has that :

(i) : the multiplication is an analytic mapping from $\mathcal{A}\times 
\mathcal{A}$ onto $\mathcal{A}$.

(ii) : the involution $*$ is a $C^{\infty }$-diffeomorphism of $\mathcal{A}$.
\end{proposition}

\proof%
%
Let $m:\mathcal{A}\times \mathcal{A}\rightarrow \mathcal{A}$, defined by $%
m(a,b)=a.b$, $(a,b)\in \mathcal{A}\times \mathcal{A}$, and let $\theta :%
\mathcal{A}\rightarrow \mathcal{A}$ defined by $\theta (a)=a^{*},a\in 
\mathcal{A}$. The $\mathcal{K}$-topology of $\mathcal{A}$ being given by the
norms $\left( \left\| .\right\| _{n}\right) _{n\in \mathbb{N}}$, by Lemma
3.1 one has that $m$ is continuous on $\mathcal{A}\times \mathcal{A}$ and
that $\theta $ is continuous on $\mathcal{A}$. Assertion (i) follows then
from the fact that $m$ is a continuous bilinear mapping (see \S\ 2.1).

Assertion (ii) follows from the continuity of $*=\theta =\theta ^{-1}$, and
from the fact that given any pair $(x,v)$ of elements in $\mathcal{A}\times 
\mathcal{A}$, by a trivial computation one checks that the corresponding
G\^{a}teaux derivative is :

\begin{equation*}
\theta ^{\prime }(x;v)=\underset{t\rightarrow 0}{\lim }\frac{\theta
(x+tv)-\theta (x)}{t}=v^{*}=\theta (v),
\end{equation*}
so that : $v\in \mathcal{A}\mapsto \theta ^{\prime }(x;v)=\theta (v)\in 
\mathcal{A}$ is a continuous complex linear mapping from $\mathcal{A}$ into $%
\mathcal{A}$, and then (see \S\ 1.1) one has that $\theta =\theta ^{-1}$ is $%
C^{\infty }$%
\endproof%
%

\subsection{Exponential mappings for a good spectral triple}

a) Let $\mathcal{K}=(\mathcal{A},\mathcal{H},D)$ be a good spectral triple.
By Proposition 3.3, the family $\{\mathcal{A}_{k},\left\| .\right\|
_{k},i_{k},k\in \mathbb{N}\}$ is a projective system of unital Banach $*$%
-algebras. One deduces that the sheaf of (germs of) analytic functions on $%
\mathcal{A}$, i.e. the projective limit of sheaves of (germs of) analytic
functions on the Banach spaces $\mathcal{A}_{k}$, is the sheaf of (germs of)
restrictions to $\mathcal{A}$ of analytic functions on the $\mathcal{A}_{k}$%
, $k\in \mathbb{N}$. Moreover, considering $\mathcal{A}$ and the $\mathcal{A}%
_{k}$, $k\in \mathbb{N}$, with their natural structure of Lie algebra given
the canonical Lie bracket $[a,b]=a.b-b.a$, one has that $\mathcal{A}$ is a
Fr\'{e}chet Lie algebra and that all the $\mathcal{A}_{k}$, $k\in \mathbb{N}$%
, are Banach Lie algebras.\\

b) Let us recall now well known results related to unital Banach algebras
and that we summarize here (see e.g. Theorem 22A in [7], \S\ 7 in Chap. II
of [4], or \S\ II.7 in [12]) : given an unital Banach algebra $\mathcal{B}$
over $\mathbb{C}$, the group $Inv(\mathcal{B})$ of its invertible elements
has a natural structure of analytic Banach Lie group open in $\mathcal{B}$,
with Lie algebra $\mathcal{B}$, and with exponential mapping :

\begin{equation*}
\exp :b\in \mathcal{B}\longmapsto \exp v=\underset{n=0}{\overset{n=\infty }{%
\sum }}\frac{v^{n}}{n!}\in Inv(\mathcal{B}),
\end{equation*}
which restricts into an analytic diffeomorphism from some convex open
neighborhood of 0 in the algebra onto an open neighborhood of $\mathbb{I}$
in $Inv(\mathcal{B})$.

Furthermore, one has that any continuous 1-parameter subgroup of $Inv(%
\mathcal{B})$ is of the form $t\mapsto \exp (tX)$ for some $X$ in $\mathcal{B%
}$. We have also to observe that $(Inv(\mathcal{B}),\mathcal{B},\exp )$ is
of course a generalized Lie group.\\

c) Let us denote by $Inv(\mathcal{A})$ the group of invertible elements in $%
\mathcal{A}$ and by $Inv(\mathcal{A}_{k})$ the group of invertible elements
in $\mathcal{A}_{k}$, $k\in \mathbb{N}$, and let us consider now the
exponential series $Exp$ of $\mathcal{A}$ (resp. : $Exp_{k}$ of $\mathcal{A}%
_{k}$, $k\in \mathbb{N}$) respectively defined by :

\begin{equation}
Exp(v)=\underset{n=0}{\overset{n=\infty }{\sum }}\frac{v^{n}}{n!},v\in 
\mathcal{A}\text{ and }Exp_{k}(v)=\underset{n=0}{\overset{n=\infty }{\sum }}%
\frac{v^{n}}{n!},v\in \mathcal{A}_{k},k\in \mathbb{N}.  \tag{4}  \label{4}
\end{equation}

As a consequence of the above discussion one obtains :

\begin{lemma}
Let $\mathcal{K}=(\mathcal{A},\mathcal{H},D)$ be a good spectral triple.
Then, for any element $k$ in $\mathbb{N}$, the group $Inv(\mathcal{A}_{k})$
has a natural structure of analytic Banach Lie group with Lie algebra $%
\mathcal{A}_{k}$, and with exponential mapping $Exp_{k}$ defined by (4) ; in
particular $(Inv(\mathcal{A}_{k}),\mathcal{A}_{k},Exp_{k})$ is a generalized
Lie group. Moreover :

(1) : $Inv(\mathcal{A}_{k})$ is open in $\mathcal{A}_{k}$ ;

(2) : there exists an open neighborhood $V_{k}$ of 0 in $\mathcal{A}_{k}$
and an open neighborhood $W_{k}$ of $\mathbb{I}$\textbf{\ }in $Inv(\mathcal{A%
}_{k})$ such $Exp_{k}$ restricts into an analytic diffeomorphism from $V_{k}$
onto $W_{k}$ ;

(3) : $Exp$ is the restriction of $Exp_{k}$ to $\mathcal{A}$ and maps $%
\mathcal{A}$ into $Inv(\mathcal{A})$.
\end{lemma}

\subsection{On some properties of the exponential mapping}

As concerns the properties of the mapping $Exp$, $\forall v\in \mathcal{A}$, 
$\forall k\in \mathbb{N}$ and $\forall p\in \mathbb{N}-\{0\}$ one has :

\begin{equation*}
\left\| S_{n+p}(v)-S_{n}(v)\right\| \leq \frac{(\left\| v\right\| _{k})^{n+1}%
}{(n+1)!}+...+\frac{(\left\| v\right\| _{k})^{n+p}}{(n+p)!}
\end{equation*}
where for each $n\geq 0:$

\begin{equation*}
S_{n}(v)=1+v+...+\frac{v^{j}}{j!}+...+\frac{v^{n}}{n!}
\end{equation*}
is the $(n+1)$-th partial sum of $Exp(v)$. Since $\mathcal{A}$ is complete
for the $\mathcal{K}$-topology, one deduces that the exponential series $%
Exp(v)$ is uniformly and absolutely convergent on any bounded subset of $%
\mathcal{A}$ with respect to the norm $\left\| .\right\| _{k}$ for any
integer $k\geq 0$. It follows that the mapping $Exp:v\mapsto Exp(v)$ is
continuous on $\mathcal{A}$, and of course, takes its values in the group $%
Inv(\mathcal{A})$.

\begin{lemma}
Let $\mathcal{K}=(\mathcal{A},\mathcal{H},D)$ be a good spectral triple.
Considered as a mapping from $\mathcal{A}$ into $\mathcal{A}$, the mapping $%
Exp$ is analytic and is the unique analytic solution of the differential
equation : 
\begin{equation*}
\frac{d}{dt}\left( \psi (tx)\right) =x.\psi (tx),\quad \psi (0x)=\mathbb{I}%
\mathbf{,\quad }x\in \mathcal{A}.
\end{equation*}
\end{lemma}

\proof%
%
$\forall t\in \mathbb{R}$ and $\forall n\in \mathbb{N}^{*}$, $\forall
(v,w)\in \mathcal{A}^{2}$ a trivial computation shows that :

\begin{equation*}
\frac{(v+tw)^{n}-v^{n}}{n!}%
=tF_{n}^{1}(v,w)+t^{2}F_{n}^{2}(v,w)+...+t^{n}F_{n}^{n}(v,w),
\end{equation*}

where, for any integer $d$ such that $1\leq d\leq n$, $F_{n}^{d}(v,w)$ is an
homogeneous polynomial function with total degree $n$. One obtains in
particular that :

\begin{equation*}
\underset{t\rightarrow 0}{Lim}\frac{(v+tw)^{n}-v^{n}}{tn!}=F_{n}^{1}(v,w)=%
\frac{1}{n!}\underset{j=0}{\overset{j=n-1}{\sum }}v^{j}.w.v^{n-1-j},
\end{equation*}
and then, for any integer $k$ :

\begin{equation}
\left\| F_{n}^{1}(v,w)\right\| _{k}\leq \frac{\left\| w\right\| _{k}}{n!}%
\underset{j=0}{\overset{j=n-1}{\sum }}(\left\| v\right\| _{k})^{n-1}=\left\|
w\right\| _{k}\frac{(\left\| v\right\| _{k})^{n-1}}{(n-1)!}.  \tag{5}
\label{5}
\end{equation}

Let us consider now the series :

\begin{equation*}
\underset{n=0}{\overset{n=\infty }{\sum }}F_{n}^{1}(v,w)=\underset{n=0}{%
\overset{\infty }{\sum }}\underset{t\rightarrow 0}{Lim}\frac{(v+tw)^{n}-v^{n}%
}{tn!}.
\end{equation*}

By (4) one has :

\begin{equation*}
\left\| \underset{n=0}{\overset{\infty }{\sum }}F_{n}^{1}(v,w)\right\|
_{k}\leq \left\| w\right\| _{k}\exp (\left\| w\right\| _{k}),k\in \mathbb{N}.
\end{equation*}

It follows that the exponential mapping $Exp$ has a G\^{a}teaux derivative
at any point $v$ in $\mathcal{A}$ in any direction $w$ of $\mathcal{A}$,
given by :

\begin{equation}
Exp^{\prime }(v;w)=\underset{n=0}{\overset{n=\infty }{\sum }}F_{n}^{1}(v,w)=%
\underset{n=0}{\overset{n=\infty }{\sum }}\frac{1}{n!}\underset{n=0}{%
\overset{n=\infty }{\sum }}v^{j}.w.v^{n-1-j}.  \tag{6}  \label{6}
\end{equation}

One deduces that $Exp^{\prime }:(v,w)\in \mathcal{A}\times \mathcal{A}%
\mapsto Exp_{v}^{\prime }(w)=Exp^{\prime }(v;w)\in \mathcal{A}$ is
continuous, and that $Exp_{v}^{\prime }:w\in \mathcal{A}\mapsto
Exp_{v}^{\prime }(w)=Exp^{\prime }(v;w)\in \mathcal{A}$ is a $\mathbb{C}$%
-linear mapping.

According to the notion of analyticity recalled in 1.1, one deduces that $%
Exp $ is a complex analytic mapping from $\mathcal{A}$ into $\mathcal{A}$
taking its values in $Inv(\mathcal{A})$.

Now, considering for each element $v$ in $\mathcal{A}$ the mapping $\psi
_{v}:\mathbb{R\rightarrow }\mathcal{A}$ defined by $\psi _{v}(t)=Exp(tv)$
which takes its values in $Inv(\mathcal{A})$, a trivial computation gives :

\begin{equation*}
\frac{d}{dt}(\psi _{v}(t))=v.\psi _{v}(t)\text{, with }\psi _{v}(0)=\mathbb{I%
}.
\end{equation*}

The assertion follows from the fact that the equation $\frac{d}{dt}(\psi
_{v}(t))=v.\psi _{v}(t)$, $\psi _{v}(0)=\mathbb{I}$, $x\in \mathcal{A}$, has
at most one analytic solution (Cf. [8], Assertion 3.11)%
\endproof%
%

From Lemmas 4.4 and 4.5 one deduces easily that :

\begin{lemma}
Let $\mathcal{K}=(\mathcal{A},\mathcal{H},D)$ be a good spectral triple ;
then $(Inv(\mathcal{A}),\mathcal{A},Exp)$ is a generalized Lie group.
\end{lemma}

\section{The good unital $*$-algebra of a good spectral triple}

\subsection{On the induced $\mathcal{K}$-topology}

Let $\mathcal{K}=(\mathcal{A},\mathcal{H},D)$ be a good spectral triple. In
2.2(a) it has be shown that, for any element $n$ in $\mathbb{N}$, the
natural embedding $i_{n}:\mathcal{A}_{n+1}\hookrightarrow \mathcal{A}_{n}$
is a continuous morphism of Banach $*$-algebras with dense range, so that $\{%
\mathcal{A}_{n},\left\| .\right\| _{n},i_{n},n\in \mathbb{N}\}$ is a
projective system of unital involutive Banach algebras, the projective limit
of which is the unital Fr\'{e}chet $*$-algebra :

\begin{equation*}
\mathcal{A}=\underset{\longleftarrow }{Lim}\mathcal{A}_{n}=\underset{n\geq 0%
}{\bigcap }\mathcal{A}_{n}.
\end{equation*}

At the level of the corresponding Banach Lie groups $Inv(\mathcal{A}_{n})$, $%
n\in \mathbb{N}$, one has to observe that $i_{n}(Inv(\mathcal{A}_{n+1}))$ is
a subgroup of $Inv(\mathcal{A}_{n})$, that the restriction $j_{n}$ of $i_{n}$
to $Inv(\mathcal{A}_{n+1})$ is a continuous morphism of topological groups,
and that :

\begin{equation}
\left\{ 
\begin{array}{l}
Inv(\mathcal{A}_{n+1})=Inv(\mathcal{A}_{n})\bigcap Inv(\mathcal{A}%
_{n+1}=(j_{n})^{-1}(Inv(\mathcal{A}_{n})) \\ 
j_{n}\circ \exp _{n+1}=\exp _{n}\circ i_{n}
\end{array}
\right\}  \tag{7}  \label{7}
\end{equation}

Endly, the embedding $i_{n}:\mathcal{A}_{n+1}\hookrightarrow \mathcal{A}_{n}$
being in particular a continuous linear mapping, is analytic (see 1.1), and
since $Inv(\mathcal{A}_{n+1})$ and $Inv(\mathcal{A}_{n})$ are open in
respectively $\mathcal{A}_{n+1}$ and $\mathcal{A}_{n}$, one has that $j_{n}$
is an analytic injection of Banach Lie groups from $Inv(\mathcal{A}_{n+1})$
into $Inv(\mathcal{A}_{n})$.

\begin{lemma}
Let $\mathcal{K}=(\mathcal{A},\mathcal{H},D)$ be a good spectral triple.
Provided with the topology induced from the $\mathcal{K}$-topology of $%
\mathcal{A}$ the group $Inv(\mathcal{A})$ is a topological group.
\end{lemma}

\proof%
%
Let $\mathcal{K}=(\mathcal{A},\mathcal{H},D)$ be a good spectral triple, and
let us provide the group $Inv(\mathcal{A})$ with induced topology of the $%
\mathcal{K}$-topology of $\mathcal{A}$.

It follows from the above discussion that it is the weakest topology for
which all the embedding of typological groups :

\begin{equation*}
j_{n}:Inv(\mathcal{A}_{n+1})\hookrightarrow Inv(\mathcal{A}_{n})
\end{equation*}
are continuous, so that $\{Inv(\mathcal{A}_{n}),j_{n},n\in \mathbb{N}\}$ is
a projective system of Banach Lie groups, having as projective limit $Inv(%
\mathcal{A})$ provided with the induced $\mathcal{K}$-topology. Moreover,
one has that $\{\mathcal{A}_{n},\left\| .\right\| _{n},i_{n},n\in \mathbb{N}%
\}$ into the projective system of topological spaces $\{Inv(\mathcal{A}%
_{n}),j_{n},n\in \mathbb{N}\}$, so that $Exp=\underset{\longleftarrow }{Lim}%
Exp_{n}$.

For any $n$ in $\mathbb{N}$, let $m_{n}:Inv(\mathcal{A}_{n})\times Inv(%
\mathcal{A}_{n})\rightarrow Inv(\mathcal{A}_{n})$ be the multiplication on $%
Inv(\mathcal{A}_{n})$, and let $inv_{n}:Inv(\mathcal{A}_{n})\rightarrow Inv(%
\mathcal{A}_{n})$ be the mapping :

\begin{equation*}
a\in Inv(\mathcal{A}_{n})\mapsto Inv(a)=a^{-1}\in Inv(\mathcal{A}_{n}).
\end{equation*}

One easily sees that $\{m_{n},n\in \mathbb{N}\}$ is a projective system of
smooth mappings from the projective system of Banach Lie groups $\{Inv(%
\mathcal{A}_{n})\times Inv(\mathcal{A}_{n}),j_{n},n\in \mathbb{N}\}$ into
the projective system of Banach Lie groups $\{Inv(\mathcal{A}%
_{n}),j_{n},n\in \mathbb{N}\}$, and that $\{inv_{n},n\in \mathbb{N}\}$ is a
projective system of smooth mappings from the projective system of Banach
Lie groups $\{Inv(\mathcal{A}_{n}),j_{n},n\in \mathbb{N}\}$ into itself. The
proof follows them from the fact that the multiplication :

\begin{equation*}
m=\underset{\longleftarrow }{\lim }m_{n}:Inv(\mathcal{A}_{n})\times Inv(%
\mathcal{A}_{n})\rightarrow Inv(\mathcal{A}_{n})
\end{equation*}
is continuous on $Inv(\mathcal{A})\times Inv(\mathcal{A})$, and likewise,
that the mapping :

\begin{equation*}
inv=\underset{\longleftarrow }{\lim }inv_{n}:a\in Inv(\mathcal{A}%
_{n})\mapsto Inv(a)=a^{-1}\in Inv(\mathcal{A}_{n})
\end{equation*}
is continuous on $Inv(\mathcal{A})$%
\endproof%
%

\subsection{On the category of good unital Fr\'{e}chet $*$-algebras}

a) Adapting a notion initiated by R. Swan ([14]) for topological rings,
J.-B. Bost has considered in [3] the class of\textit{\ good topological
algebras}, namely : the class of unital topological associative algebras $A$
over $\mathbb{C}$ for which :

-the group $Inv(A)$ of invertible elements in $A$ open in $A$,

-the mapping $a\in Inv(A)\mapsto a^{-1}\in Inv(A)$ is continuous..

It is well known that any unital Banach algebra $A$ is a good topological
algebra since, in such a case, the group $Inv(A)$ is a Banach Lie group open
in $A$ (see e.g. [4] and [7]) ; but there exists unital Fr\'{e}chet algebras
which are not good topological algebras (see e.g. [3], A.1.2).

\begin{definition}
By good unital Fr\'{e}chet $*$-algebra, it will be meant an unital
Fr\'{e}chet $*$-algebra which, as unital topological algebra, is a good
algebra.
\end{definition}

b) As a matter of fact, the good complete Hausdorff locally convex
topological algebras constitute the objects of a category $\mathcal{G}$ on
which one can define the functors $K_{0}$ and $K_{1}$ of the topological $K$%
-theory, on account of the fact that such algebras have a holomorphic
functional calculus ([3], A.1.5). One obtains then a subcategory, by taking
as objects the good unital Fr\'{e}chet $\ast $-algebras, and as morphisms
the continuous $\ast $-morphisms of Fr\'{e}chet *-algebras.

Given two good unital Fr\'{e}chet $*$-algebras $A$ and $B$, any continuous $%
* $-morphism $\phi :A\rightarrow B$ such that $\phi (\mathbb{I}_{A})=\mathbb{%
I}_{B} $ induces in a standard way a morphism of abelian groups :

\begin{equation*}
\phi _{*}:K_{j}(A)\rightarrow K_{j}(B),j=0,1.
\end{equation*}

As a direct application of an important result of J.-B. Bost ([3], Theorem
A.1.2) to which we refer, and which generalizes a well known result of R.
Swan ([14], 2.2) related to the case of unital Banach algebras, one obtains :

\begin{lemma}
Let $A$ and $B$ be two good unital Fr\'{e}chet $*$-algebras, and let $\phi $
be a continuous unital $*$-morphism from $A$ into $B$ such that :

(ii) : $\phi (A)$ is dense in $B$,

(ii) : $\phi ^{-1}(Inv(B))=Inv(A)$.

Then $\phi _{*}:K_{i}(A)\rightarrow K_{i}(B)$, $i=0,1$, is an isomorphism of
groups.
\end{lemma}

\subsection{The main results}

\begin{theorem}
Let $\mathcal{K}=(\mathcal{A},\mathcal{H},D)$ be a good spectral triple, and
let us provide $\mathcal{A}$ with the corresponding $\mathcal{K}$-topology.
Then, $Inv(A)$ has a natural structure of analytic Fr\'{e}chet Lie group
open in its Lie algebra $\mathcal{A}$, and for any integer $k\geq 0$ one has
: $Inv(\mathcal{A}_{k})=\mathcal{A}_{k}\bigcap Inv(\mathcal{A}_{0})$ ;
moreover, its exponential mapping $Exp:\mathcal{A}\rightarrow Inv(\mathcal{A}%
)$ is analytic and restricts into an analytic diffeomorphism from some open
neighborhood of zero in $\mathcal{A}$ onto some open neighborhood of the
unit $\mathbb{I}$ in $Inv(\mathcal{A})$. In particular $\mathcal{A}$ is a
good unital Fr\'{e}chet $*$-algebra and $Inv(\mathcal{A})$ is a Lie group of
Campbell-Baker-Hausdorff type.
\end{theorem}

\proof%
%
a) Let us prove firstly that $Inv(\mathcal{A})$ is open in $\mathcal{A}$ and
that for any integer $k\geq 0$ one has :

\begin{equation*}
Inv(\mathcal{A}_{k})=\mathcal{A}_{k}\bigcap Inv(\mathcal{A}_{0}).
\end{equation*}

Taking into account that for any $k$ in $\mathbb{N}$, the Banach Lie group $%
Inv(\mathcal{A}_{k})$ is open in the Banach algebra $\mathcal{A}_{k}$, that $%
\mathcal{A}=\underset{n\geq 0}{\bigcap }\mathcal{A}_{n}$ and that $Inv(%
\mathcal{A})=\underset{n\geq 0}{\bigcap }Inv(\mathcal{A}_{n})$, it suffices
to prove that :

\begin{equation*}
Inv(\mathcal{A}_{k})=\mathcal{A}_{k}\bigcap Inv(\mathcal{A}_{0})\quad
\forall k\in \mathbb{N}.
\end{equation*}

In particular, for $k=0$, since the exponential mapping $\exp _{0}:\mathcal{A%
}_{0}\rightarrow Inv(\mathcal{A}_{0})$ restricts into an analytic
diffeomorphism from some convex open neighborhood of $0$ onto an open
neighborhood of $\mathbb{I}$ in $Inv(\mathcal{A}_{0})$, we can find a real
number $\varepsilon $, $0<\varepsilon <\frac{1}{2}$ such that :

\begin{equation*}
W=\{1+v/v\in \mathcal{A}_{0}\text{ and }\left\| v\right\| _{0}<\varepsilon \}
\end{equation*}
is an open neighborhood of $\mathbb{I}$ in $Inv(\mathcal{A}_{0})$. The fact
that $Inv(\mathcal{A})$ is open in $\mathcal{A}$ is a now consequence of the
following two assertions :

-Assertion $(P):W\bigcap \mathcal{A}_{k}\subset Inv(\mathcal{A}_{k})$, $%
\forall k\in \mathbb{N}$,

-Assertion $(Q):\mathcal{A}_{k}\bigcap Inv(\mathcal{A}_{0})\subset Inv(%
\mathcal{A}_{k})$, $\forall k\in \mathbb{N}$.

a-1) : Proof of Assertion $(P)$. Observing that this assertion is trivial
for $k=0$, let us assume that for some integer $n\geq 1:W\bigcap \mathcal{A}%
_{n-1}\subset Inv(\mathcal{A}_{n-1})$. In order to prove that $W\bigcap 
\mathcal{A}_{n}\subset Inv(\mathcal{A}_{n})$, taking into account that $%
W\bigcap Inv(\mathcal{A}_{n})$ is open in the connected topological space $%
W\bigcap \mathcal{A}_{n}$, it suffices to prove that $W\bigcap Inv(\mathcal{A%
}_{n})$ is closed in $W\bigcap \mathcal{A}_{n}$.

Given any element $a$ in the boundary of $W\bigcap Inv(\mathcal{A}_{n})$ in $%
W\bigcap \mathcal{A}_{n}$, let $(a_{p})_{p}$ be a sequence of elements in $%
W\bigcap Inv(\mathcal{A}_{n})\bigcap \mathcal{A}$ converging to $a$ with
respect to $\left\| .\right\| _{n}$ and let us consider the mapping $l(a):%
\mathcal{A}_{n}\rightarrow \mathcal{A}_{n}$ (resp : $l(a_{p}):\mathcal{A}%
_{n}\rightarrow \mathcal{A}_{n},p\in \mathbb{N}$) consisting of the left
mutiplication by $a$ (resp. : by $a_{p},p\in \mathbb{N}$).

We want to prove, first of all, that $l(a)$ is a continuous bijection on $%
\mathcal{A}_{n}$. In fact, although the $l(a_{p})$ $p\in \mathbb{N}$, are
continuous bijections on $\mathcal{A}_{n}$, we can only assert that $l(a)$
is a continuous injection, so that it remains to prove that $l(a)$ is a
surjective mapping. This proof is based on two points :

(a-1-$\alpha $) : Since $a$ lies in $W\bigcap \mathcal{A}_{n}$ for every $x$
in $\mathcal{A}_{n}$, a first point is that, by Lemma 3.2 :

\begin{equation*}
\left\| x\right\| _{n}-\left\| l(a)x\right\| _{n}=\left\| x\right\|
_{n}-\left\| a.x\right\| _{n}\leq \left\| (1-a).x\right\| _{n}\leq \frac{1}{2%
}\left\| x\right\| _{n}-\eta _{a,n}\left\| x\right\| _{n-1},
\end{equation*}
so that one obtains the inequality : $\left\| l(a)x\right\| _{n}\geq \frac{1%
}{2}\left\| x\right\| _{n}-\eta _{a,n}\left\| x\right\| _{n-1}$, which,
taking into account the inductive assumption, implies that $l(a)(\mathcal{A}%
_{n})=a.\mathcal{A}_{n}$ is closed in $\mathcal{A}_{n}$.

(a-1-$\beta $) : The second point consists of proving that $\mathcal{A}%
_{n}\subset l(a)(\mathcal{A}_{n})$, so that $l(a)$ is a bijective mapping.
Let $z$ be in $\mathcal{A}_{n}$ ; since $l(a_{p})$ is bijective for each $p$
in $\mathbb{N}$, we can find a sequence $(x_{p})_{p}$ of elements in $%
\mathcal{A}_{n}$ such that : $z=l(a_{p})x_{p}=a_{p}.x_{p}$.

Now, $a.\mathcal{A}_{n}$ being closed in $\mathcal{A}_{n}$ one has :

\begin{equation*}
\underset{p\rightarrow \infty }{\lim }a_{p}.x_{p}=\underset{p\rightarrow
\infty }{\lim }a.x_{p}
\end{equation*}
so that one has to prove that $\underset{p\rightarrow \infty }{\lim }%
a.x_{p}=z$. By Lemma 4.5, for any $p$ in $\mathbb{N}^{*}$ one has :

\begin{equation*}
\left\| a_{p}.x_{p}\right\| _{n}\geq \left\| a.x_{p}\right\| _{n}-\left\|
a_{p}-a\right\| _{n}\left\| x_{p}\right\| _{n}\geq \frac{1}{2}\left\|
x_{p}\right\| _{n}-\eta _{a,n}\left\| x_{p}\right\| _{n-1}-\left\|
a_{p}-a\right\| _{n}\left\| x_{p}\right\| _{n},
\end{equation*}
from which it follows, for sufficiently large $p$, that

\begin{equation*}
\left\| a_{p}.x_{p}\right\| _{n}\geq \varepsilon \left\| x_{p}\right\|
_{n}-\eta _{a,n}\left\| x_{p}\right\| _{n-1}.
\end{equation*}

Let us prove now that $W\bigcap Inv(\mathcal{A}_{n})$ is closed in $W\bigcap 
\mathcal{A}_{n}$ ; taking into account the inductive assumption which
implies that $\left( \left\| x_{p}\right\| _{n-1}\right) _{p}$ is a bounded
sequence, the inequality (9) implies that $\left( \left\| x_{p}\right\|
_{n}\right) _{p}$ is bounded ; then, by Lemma 3.1 :

\begin{equation*}
\left\| z-a.x_{p}\right\| _{n}=\left\| a_{p}.x_{p}-a.x_{p}\right\| _{n}\leq
\left\| (a_{p}-a).x_{p}\right\| _{n}\leq \left\| a_{p}-a\right\| _{n}\left\|
x_{p}\right\| _{n}
\end{equation*}
proving that $\underset{p\rightarrow \infty }{\lim }a.x_{p}=z$, that $a\in
Inv(\mathcal{A}_{n})$, and that $W\bigcap Inv(\mathcal{A}_{n})$ is closed in 
$W\bigcap \mathcal{A}_{n}$.

a-2) : Proof of Assertion $(Q)$. Let $x$ be any element in $\mathcal{A}%
_{k}\bigcap Inv(\mathcal{A}_{0})$. One has in particular that $x^{-1}$ lies
in the Banach Lie group $Inv(\mathcal{A}_{0})$, so that we can find a
sequence $(v_{p})_{p}$ in $\mathcal{A}$ such that $\underset{p\rightarrow
\infty }{\lim }\left\| v_{p}-x^{-1}\right\| _{0}=0$ and then, for
sufficiently large $p$ one obtains : $\left\| 1-v_{p}.x\right\|
_{0}<\varepsilon $.\\

b) We have now to prove that $Inv(\mathcal{A})$ has a structure of
Fr\'{e}chet Lie group with Lie algebra $\mathcal{A}$. First of all, since $%
Inv(\mathcal{A})$ is open in the Fr\'{e}chet algebra $\mathcal{A}$, $Inv(%
\mathcal{A})$ inherits a natural structure of Fr\'{e}chet manifold, the
underlying topology being the induced $\mathcal{K}$-topology.

Then, taking into account that $\{Inv(\mathcal{A}_{n}),j_{n},n\in \mathbb{N}%
\}$ is a projective system of analytic Banach Lie groups having $Inv(%
\mathcal{A}) $ as projective limit (see the proof of Lemma 5.1), by Lemma
4.4 one has that $\{Exp_{n},n\in \mathbb{N}\}$ is a projective system of
analytic mappings from the projective system of Banach algebras $\{\mathcal{A%
}_{n},n\in \mathbb{N}\}$ into the projective system of analytic Banach Lie
groups $\{Inv(\mathcal{A}_{n}),j_{n},n\in \mathbb{N}\}$, so that $Exp=%
\underset{\leftarrow }{\lim }Exp_{n}$ is an analytic mapping from $\mathcal{A%
}$ into $Inv(\mathcal{A})$.

We have to recall now, by Proposition 4.3 that the multiplication $m:%
\mathcal{A}\times \mathcal{A}\rightarrow \mathcal{A}$ is an analytic mapping
; $Inv(\mathcal{A})$ being open in $\mathcal{A}$, one easily deduces that
its restriction, still denoted by $m$, to $Inv(\mathcal{A})\times Inv(%
\mathcal{A})$, is analytic mapping from $Inv(\mathcal{A})\times Inv(\mathcal{%
A})$ into $Inv(\mathcal{A})$.

Moreover, denoting by $inv_{n}$ the mapping : 
\begin{equation*}
a\in Inv(\mathcal{A}_{n})\rightarrow inv_{n}(a)=a^{-1}\in Inv(\mathcal{A}%
_{n}),n\in \mathbb{N},
\end{equation*}
$\{inv_{n},n\in \mathbb{N}\}$ is a projective system of analytic mappings
from the projective system of analytic Banach Lie groups $\{Inv(\mathcal{A}%
_{n}),j_{n},n\in \mathbb{N}\}$ into itself, and then, the mapping :

\begin{equation*}
inv=\underset{\leftarrow }{\lim }inv_{n}:a\in Inv(\mathcal{A})\longmapsto
inv(a)=a^{-1}\in Inv(\mathcal{A})
\end{equation*}
is analytic on $Inv(\mathcal{A})$. Since, by Lemma 4.6, ($Inv(\mathcal{A}),%
\mathcal{A},Exp)$ is a generalized Lie group, using standard arguments, (see
e.g. [12]), one easily concludes that the group $Inv(\mathcal{A})$ provided
with the Fr\'{e}chet manifold structure inherited from that of the
Fr\'{e}chet space $\mathcal{A}$ is an analytic Fr\'{e}chet Lie group open in 
$\mathcal{A}$, with Lie algebra $\mathcal{A}$ and exponential mapping $Exp$.\\

c) It is well known that, given any analytic Banach Lie group $G$ with Lie
algebra $\mathcal{G}$ and exponential mapping $\exp :\mathcal{G}\rightarrow
G $ is a Lie group of Campbell-Baker-Hausdorff type, i.e. is such that $\exp 
$ restricts into analytic diffeomorphism from some neighborhood of zero in $%
\mathcal{G}$ into some neighborhood of $\mathbb{I}$ in $G$ (see e.g. [11],
\S 1).

Since $\{Exp_{n},n\in \mathbb{N}\}$ is a projective system of analytic
mappings from the projective system of Banach algebras $\{\mathcal{A}%
_{n},i_{n,}n\in \mathbb{N}\}$ into the projective system of analytic Banach
Lie groups $\{Inv(\mathcal{A}_{n}),j_{n},n\in \mathbb{N}\}$, one deduces
easily that exponential mapping $Exp:\mathcal{A}\rightarrow Inv(\mathcal{A})$
is analytic and restricts into an analytic diffeomorphism from some open
neighborhood of zero in $\mathcal{A}$ onto some open neighborhood of $%
\mathbb{I}$ in $Inv(\mathcal{A})$. $Inv(\mathcal{A})$ is then a Fr\'{e}chet
Lie group of Campbell-Baker-Hausdorff type and $\mathcal{A}$ is a good
unital Fr\'{e}chet $*$-algebra%
\endproof%
%

\begin{theorem}
Let $\mathcal{K}=(\mathcal{A},\mathcal{H},D)$ be a good spectral triple, and
let $\mathcal{A}_{0}$ be the completion of $\mathcal{A}$ with respect to the 
$C^{*}$-norm $\left\| .\right\| $ of operators in $\mathcal{H}$.

(1) : The natural embedding $i$ of $\mathcal{A}$ in the unital $C^{*}$%
-algebra $\mathcal{A}_{0}$ induces, for the associated $K$-theories, an
isomorphism of groups $i_{*}:K_{j}(\mathcal{A})\rightarrow K_{j}(\mathcal{A}%
_{0}),j=0,1$.

(2) : For any integer $n\geq 1$, let $I_{n}$ be the identity on $\mathbb{C}%
^{n}$ and let $M_{n}(\mathcal{A})$ be the unital $*$-algebra of $n$ by $n$
matrices over $\mathcal{A}$. Then, $\mathcal{K}_{n}=(M_{n}(\mathcal{A}),%
\mathcal{H}\mathbb{\otimes C}^{n},D\mathbb{\otimes I}_{n})$ is good spectral
triple.
\end{theorem}

\proof%
%
(1) : Since any unital $C^{*}$-algebra is a good algebra, assertion (1) is
an obvious consequence of Theorem 5.3 and Lemma 5.2.

(2) : First of all, one easily sees that $\mathcal{K}_{n}$ is exactly the
tensor product of the good spectral triple $\mathcal{K}=(\mathcal{A},%
\mathcal{H},D)$ with the trivial spectral triple $(M_{n}(\mathbb{C}),\mathbb{%
C}^{n},\mathbb{I}_{n})$ (see [5]), so that $\mathcal{K}_{n}$ is a spectral
triple. Moreover, since for all integer $p\geq 0$ the algebra $\mathcal{A}$
is contained in the domain of the $p$-th power of the derivation $[\left|
D\right| ,.]$, one deduces that $M_{n}(\mathcal{A})$ lies in the domain of
the $p$-th power of the derivation $[\left| D\otimes \mathbb{I}_{n}\right|
,.]$ so that $\mathcal{K}_{n}$ is a regular spectral triple. Endly, since $\{%
\mathcal{A},\mathcal{A}_{k};k\in \mathbb{N}\}$ is an ILB-chain, one easily
check that $\{M_{n}(\mathcal{A})(\mathcal{K}_{n}),M_{n}(\mathcal{A}%
_{k});k\in \mathbb{N}\}$ is an ILB-chain, from which one deduces that :

\begin{equation*}
M_{n}(\mathcal{A})(\mathcal{K}_{n})=\underset{\longleftarrow }{\lim }M_{n}(%
\mathcal{A}_{k})=\underset{k\geq 0}{\bigcap }M_{n}(\mathcal{A}_{k})=M_{n}(%
\underset{k\geq 0}{\bigcap }\mathcal{A}_{k})=M_{n}(\mathcal{A})
\end{equation*}
so that, finally, $\mathcal{K}_{n}$ is a good spectral triple and that $%
M_{n}(\mathcal{A})$ is a good unital Fr\'{e}chet $*$-algebra. Furthermore,
the equality :

\begin{equation*}
\underset{k\geq 0}{\bigcap }M_{n}(\mathcal{A}_{k})=M_{n}(\underset{k\geq 0}{%
\bigcap }\mathcal{A}_{k})=M_{n}(\mathcal{A})
\end{equation*}
implies that the $\mathcal{K}_{n}$-topology is the $n^{2}$-fold product of
the $\mathcal{K}$-topology of $\mathcal{A}$%
\endproof%
%

\begin{theorem}
Let $\mathcal{K}=(\mathcal{A},\mathcal{H},D)$ be a good spectral triple and
let $n$ be any positive integer. Then :

(i) : The group $Inv(M_{n}(\mathcal{A}))$ of invertible elements of $M_{n}(%
\mathcal{A})$ has a natural structure of Fr\'{e}chet Lie group of
Campbell-Baker-Hausdorff type with Lie algebra $M_{n}(\mathcal{A})$ and
exponential mapping $Exp_{n}:M_{n}(\mathcal{A})\longrightarrow Inv(M_{n}(%
\mathcal{A}))$ given for any element $v$ in $M_{n}(\mathcal{A})$ by : 
\begin{equation*}
Exp_{n}(v)=\underset{n\geq 0}{\sum }\frac{v^{n}}{n!}.
\end{equation*}

(ii) : Given any closed Lie subalgebra $\mathcal{L}$ of $M_{n}(\mathcal{A})$%
, there exists an unique connected Lie subgroup $L$ of
Campbell-Baker-Hausdorff type of $Inv(M_{n}(\mathcal{A}))$ the Lie algebra
of which is $\mathcal{L}$.
\end{theorem}

\proof%
%
Assertion (i) is a consequence of Theorems 5.3 and 5.4, Formula (4) and
Lemma 4.4. Assertion (ii) follows from (i) and from Theorem 3.2 of [13] (see
also [6]) which asserts that any closed Lie subalgebra of a Lie group $G$ of
Campbell-Baker-Hausdorff type is the Lie algebra of an unique connected Lie
group of Campbell-Baker-Hausdorff type embedded in $G$%
\endproof%
%

\section{On some Fr\'{e}chet Lie subgroups of $Inv(M_{n}(\mathcal{A}))$}

Let $\mathcal{K}=(\mathcal{A},\mathcal{H},D)$ be a good spectral triple and $%
n$ be any strictly positive integer.

\subsection{The Fr\'{e}chet Lie groups of Campbell-Baker-Hausdorff type $%
U_{\Omega }(n;\mathcal{A})$}

a) For any integer $n>0$, we shall consider the sets :

\begin{equation}
\mathcal{J}(n;\mathcal{A})=\{\Omega \in Inv(M_{n}(\mathcal{A}))/\Omega
^{-1}=\Omega ^{*}=\varepsilon \Omega ,\varepsilon \in \{-1,1\}\},  \tag{8}
\label{8}
\end{equation}
\begin{equation}
\mathcal{J}^{\pm }(n;\mathcal{A})=\{\Omega \in Inv(M_{n}(\mathcal{A}%
))/\Omega ^{-1}=\Omega ^{*}=\pm \Omega \},  \tag{9}  \label{9}
\end{equation}
so that $\mathcal{J}(n;\mathcal{A})=\mathcal{J}^{+}(n;\mathcal{A})\cup 
\mathcal{J}^{-}(n;\mathcal{A})$. With any element $\Omega $ in $\mathcal{J}%
^{\pm }(n;\mathcal{A})$, we associate : the mapping $\sigma _{\Omega }:M_{n}(%
\mathcal{A})\longrightarrow M_{n}(\mathcal{A})$ given by :

\begin{equation}
\sigma _{\Omega }(a)=\mp \Omega .a^{*}.\Omega ,a\in M_{n}(\mathcal{A}), 
\tag{10}  \label{10}
\end{equation}
and the sets :

\begin{equation*}
\left\{ 
\begin{array}{c}
\mathcal{U}_{\Omega }(n;\mathcal{A}))=\{a\in M_{n}(\mathcal{A})/a^{*}.\Omega
+\Omega .a=0\} \\ 
U_{\Omega }(n;\mathcal{A}))=\{a\in Inv(M_{n}(\mathcal{A}))/a^{*}.\Omega
.a=\Omega \}
\end{array}
\right\}
\end{equation*}

\begin{theorem}
Let $\mathcal{K}=(\mathcal{A},\mathcal{H},D)$ be a good spectral triple.
Given any strictly positive integer $n$ and any element $\Omega $ in $%
\mathcal{J}(n;\mathcal{A})$ one has that :

(i) : $\sigma _{J}$ is a continuous involution of the Lie algebra $M_{n}(%
\mathcal{A})$ commuting with the involution $*$.

(ii) : $\mathcal{U}_{\Omega }(n;\mathcal{A}))$ is a closed real form of the
Fr\'{e}chet Lie algebra $M_{n}(\mathcal{A})$.

(iii) : $U_{\Omega }(n;\mathcal{A}))$ is a closed Lie subgroup of
Campbell-Baker-Hausdorff type of the Fr\'{e}chet Lie group $Inv(M_{n}(%
\mathcal{A}))$, with Fr\'{e}chet Lie algebra $\mathcal{U}_{\Omega }(n;%
\mathcal{A}))$ and with exponential mapping $Exp_{\Omega }$ which is the
restriction to $\mathcal{U}_{\Omega }(n;\mathcal{A}))$ of the exponential
mapping $Exp:M_{n}(\mathcal{A})\rightarrow Inv(M_{n}(\mathcal{A}))$.
\end{theorem}

\begin{proof}
Let $\Omega $ be any element in $\mathcal{J}(n;\mathcal{A})$ so that $\Omega
^{-1}=\Omega ^{*}=\varepsilon \Omega $, with $\varepsilon =\pm 1$.\\

(i) : For every $x,y$ in $M_{n}(\mathcal{A})$ and any complex number $%
\lambda $ one has : 
\begin{equation*}
\left. 
\begin{array}{l}
\bullet :\sigma _{\Omega }(x+\lambda y)=-\varepsilon \Omega .(x+\lambda
y)^{*}.\Omega =-\varepsilon \Omega .(x^{*}+\bar{\lambda}y^{*}).\Omega
=\sigma _{\Omega }(x)+\bar{\lambda}\sigma _{\Omega }(y); \\ 
\bullet :\sigma _{\Omega }(x)^{*}=(-\varepsilon \Omega .x^{*}.\Omega
)^{*}=-\varepsilon \Omega ^{*}.(x^{*})^{*}.\Omega ^{*}=-\varepsilon
(\varepsilon \Omega ).(x^{*})^{*}.(\varepsilon \Omega )=-\varepsilon \Omega
.(x^{*})^{*}.\Omega =\sigma _{\Omega }(x^{*})
\end{array}
\right. 
\end{equation*}
which proves that $\sigma _{\Omega }$ commutes with $*$ in $M_{n}(\mathcal{A}%
)$ ; one has also : 
\begin{equation*}
\sigma _{\Omega }(\sigma _{\Omega }(x))
\begin{array}[t]{l}
=-\varepsilon \Omega .\sigma _{\Omega }(x)^{*}.\Omega  \\ 
=-\varepsilon \Omega .\sigma _{\Omega }(x^{*}).\Omega  \\ 
=-\varepsilon \Omega .(-\varepsilon \Omega .x.\Omega ).\Omega  \\ 
=(-\varepsilon \Omega ^{2}).x.(-\varepsilon \Omega ^{2}) \\ 
=x
\end{array}
\end{equation*}
which proves that $\sigma _{\Omega }$ is an antiautomorphism of order 2 of $%
M_{n}(\mathcal{A})$ ; moreover one has : 
\begin{equation*}
\sigma _{\Omega }(xy)=\sigma _{\Omega }(x).\sigma _{\Omega }(x)\text{ and }%
\sigma _{\Omega }([x,y])=[\sigma _{\Omega }(x),\sigma _{\Omega }(y)].
\end{equation*}

Endly, the continuity of $\sigma _{\Omega }$ follows from that of the
multiplication and of the involution $*$.\\

(ii) : Let us observe now, that given any element $x$ in $M_{n}(\mathcal{A})$%
, one has : 
\begin{equation*}
\sigma _{\Omega }(x)=x\Leftrightarrow x+\varepsilon \Omega .x^{*}.\Omega
=0\Leftrightarrow \Omega .(x+\varepsilon \Omega .x^{*}.\Omega
)=0\Leftrightarrow \Omega .x+\varepsilon \Omega ^{2}.x^{*}.\Omega =0
\end{equation*}
and then :

\begin{equation*}
\sigma _{\Omega }(x)=x\Leftrightarrow \Omega .x+x^{*}.\Omega
=0\Leftrightarrow x\in \mathcal{U}_{\Omega }(n;\mathcal{A})),
\end{equation*}
so that :

\begin{equation*}
\mathcal{U}_{\Omega }(n;\mathcal{A}))=\{x\in M_{n}(\mathcal{A})/x^{*}.\Omega
+\Omega .x=0\}=\{x\in M_{n}(\mathcal{A})/\sigma _{\Omega }(x)=x\}.
\end{equation*}

The equality : $\sigma _{\Omega }(x+\lambda y)=\sigma _{\Omega }(x)+%
\overline{\lambda }\sigma _{\Omega }(y),(x,y)\in M_{n}(\mathcal{A})\times
M_{n}(\mathcal{A})$ stated in (i) and the fact that $\sigma _{\Omega }$ is
an antiautomorphism of Lie algebra on $M_{n}(\mathcal{A})$ imply that $%
\mathcal{U}_{\Omega }(n;\mathcal{A}))$ is a real Lie subalgebra of $M_{n}(%
\mathcal{A})$.

By proposition 4.2 and Theorem 5.4(ii), $M_{n}(\mathcal{A})$, provided with
the $\mathcal{K}_{n}$-topology, is a Fr\'{e}chet Lie algebra. Observing that 
$\mathcal{U}_{\Omega }(n;\mathcal{A}))=\ker (\sigma _{\Omega }-Id_{M_{n}(%
\mathcal{A})})$, where $Id_{M_{n}(\mathcal{A})}$ denotes the identity on $%
M_{n}(\mathcal{A})$, the continuity of $\sigma _{\Omega }\,$implies the
closeness of $\mathcal{U}_{\Omega }(n;\mathcal{A}))$ in the Fr\'{e}chet Lie
algebra $M_{n}(\mathcal{A})$, which is then a real Fr\'{e}chet Lie
subalgebra of $M_{n}(\mathcal{A})$. Moreover, on has the direct sum : 
\begin{equation*}
M_{n}(\mathcal{A})=\mathcal{U}_{\Omega }(n;\mathcal{A}))\oplus i\mathcal{U}%
_{\Omega }(n;\mathcal{A})),
\end{equation*}
the decomposition being given by : 
\begin{equation*}
x=\frac{1}{2}(x+\sigma _{\Omega }(x))+i(\frac{1}{2}\sigma _{\Omega }(x)-ix).
\end{equation*}

$\mathcal{U}_{\Omega }(n;\mathcal{A}))$ is then a real form of $M_{n}(%
\mathcal{A})$.\\

(iii-a). Let us prove firstly that $U_{\Omega }(n;\mathcal{A}))$ is a closed
subgroup of $Inv(M_{n}(\mathcal{A}))$.

Let $a,b$ in $U_{\Omega }(n;\mathcal{A}))$ ; one has $a^{*}.\Omega .a=\Omega
=b^{*}.\Omega .b$, and then :

\begin{equation*}
(a.b)^{*}.\Omega .(ab)=b^{*}.(a^{*}.\Omega .a).b=a.\Omega .a^{*}=\Omega .
\end{equation*}

Furthermore : 
\begin{equation*}
(\Omega ^{*}.a^{*}.\Omega ).a=a.(\Omega ^{*}.a^{*}.\Omega )=\Bbb{I}_{n},
\end{equation*}
where $\Bbb{I}_{n}\Bbb{\ }$denotes the unit element of $M_{n}(\mathcal{A})$.

One deduces that any $a$ in $U_{\Omega }(n;\mathcal{A}))$ is invertible in $%
U_{\Omega }(n;\mathcal{A})$, that : 
\begin{equation*}
a\in U_{\Omega }(n;\mathcal{A})\Leftrightarrow a\in Inv(M_{n}(\mathcal{A}))%
\text{ and }a^{-1}=\Omega ^{*}.a^{*}.\Omega =\varepsilon \Omega
.a^{*}.\Omega ,
\end{equation*}
and that $U_{\Omega }(n;\mathcal{A}))$ is really a subgroup of $Inv(M_{n}(%
\mathcal{A}))$.

Endly, from the continuity of the multiplication and of the involution $*$,
one easily deduces the closeness of $U_{\Omega }(n;\mathcal{A})$ in $M_{n}(%
\mathcal{A})\,$and then in the open Fr\'{e}chet Lie group $Inv(M_{n}(%
\mathcal{A}))$.\\

(iii-b). Let $Exp_{\Omega }$ be the restriction of the exponential mapping $%
Exp_{(n)}$ from $M_{n}(\mathcal{A})$ into $Inv(M_{n}(\mathcal{A}))$ (see
Theorem 5.5) to the Lie subalgebra $\mathcal{U}_{\Omega }(n;\mathcal{A}))$
of $M_{n}(\mathcal{A}))$, let $v$ be any element in $\mathcal{U}_{\Omega }(n;%
\mathcal{A}))$ and let us compute $(Exp_{\Omega }v)^{*}.\Omega .Exp_{\Omega
}v$.

Taking into account that for any $x$ in $\mathcal{U}_{\Omega }(n;\mathcal{A}%
))$ one has $x=-\varepsilon \Omega .x^{*}.\Omega $, one obtains : 
\begin{equation*}
(Exp_{\Omega }v^{*}.\Omega .Exp_{\Omega }v.\Omega )=Exp_{\Omega
}v^{*}.\Omega .Exp_{\Omega }v.\Omega =Exp_{(n)}v^{*}.\Omega .\left( 
\underset{n\geq 0}{\sum }(-\varepsilon )^{n}\frac{(\Omega .v^{*}.\Omega )^{n}%
}{n!}\right) .\Omega .
\end{equation*}

Moreover, since $\Omega ^{2}=\varepsilon \Bbb{I}_{n}$, one has : $(\Omega
.v^{*}.\Omega )^{n}=\varepsilon ^{n-1}\Omega .(v^{*})^{n}.\Omega $, and then
: 
\begin{equation*}
(Exp_{\Omega }v)^{*}.\Omega .Exp_{\Omega }v.\Omega 
\begin{array}[t]{l}
=Exp_{(n)}v^{*}.\Omega .\left( \underset{n\geq 0}{\sum }(-1)^{n}(\varepsilon
)^{n}(\varepsilon )^{n-1}\frac{(\Omega .v^{*}.\Omega )^{n}}{n!}\right)
.\Omega \\ 
=Exp_{(n)}v^{*}.\Omega ^{2}.\left( \underset{n\geq 0}{\sum }%
(-1)^{n}\varepsilon \frac{(v^{*})^{n}}{n!}.\Omega ^{2}\right) \\ 
=\varepsilon (Exp_{(n)}v^{*}).\varepsilon \Bbb{I}_{n}.(Exp_{(n)}(-v^{*}))%
\varepsilon \varepsilon \Bbb{I}_{n}=\Bbb{I}_{n}
\end{array}
\end{equation*}

It follows that : 
\begin{equation*}
(Exp_{\Omega }v)^{*}.\Omega .Exp_{\Omega }v
\begin{array}[t]{l}
=(Exp_{\Omega }v)^{*}.\Omega .Exp_{\Omega }v.\Omega .(\varepsilon \Omega )
\\ 
=\varepsilon ^{2}\Bbb{I}_{n}.\Omega  \\ 
=\Omega 
\end{array}
\end{equation*}
and then that $Exp_{\Omega }v$ lies in $U_{\Omega }(n;\mathcal{A}))$ for any
element $v$ in $\mathcal{U}_{\Omega }(n;\mathcal{A}))$. One deduces that : 
\begin{equation*}
\mathcal{U}_{\Omega }(n;\mathcal{A}))=\{v\in \mathcal{A}/Exp_{(n)}(tv)\in
U_{\Omega }(n;\mathcal{A})\forall t\in \Bbb{R}\}.
\end{equation*}\\

(iii-c) : Since $U_{\Omega }(n;\mathcal{A}))$ is closed in $Inv(M_{n}(%
\mathcal{A}))$ and $\mathcal{U}_{\Omega }(n;\mathcal{A}))$ closed in $M_{n}(%
\mathcal{A})$, it follows from Proposition 3.4 in \cite{12} that $\left(
U_{\Omega }(n;\mathcal{A})),\mathcal{U}_{\Omega }(n;\mathcal{A}%
)),Exp_{\Omega }\right) $ is a generalized Lie group. One deduces then from
Theorem 5.5 that $\mathcal{U}_{\Omega }(n;\mathcal{A}))$ is the Lie algebra
of a unique connected Lie group $\Gamma $ of Campbell-Baker-Hausdorff type
embedded in $Inv(M_{n}(\mathcal{A}))$ with exponential mapping $Exp_{\Omega }
$. We achieve the proof by observing that, taking into account the above
discussion, $U_{\Omega }(n;\mathcal{A}))$ is a closed Fr\'{e}chet Lie
subgroup of $Inv(M_{n}(\mathcal{A}))\,$and that $\Gamma $ is necessarily the
connected component of $U_{\Omega }(n;\mathcal{A}))$%
\end{proof}%

\subsection{Examples}

Let $\mathcal{K}=(\mathcal{A},\mathcal{H},D)$ be a good spectral triple.

a) \textit{The unitary groups }$U(n;\mathcal{A})$. For any strictly positive
integer $n$ the unit element $\mathbb{I}_{n}$ of $M_{n}(\mathcal{A})$ lies
in the set $\mathcal{J}^{+}(n;\mathcal{A})$. The Fr\'{e}chet Lie group of
Campbell-Baker-Hausdorff type $U_{\mathbb{I}_{n}}(n;\mathcal{A})$, that we
shall denote by $U(n;\mathcal{A})$, is called the unitary group of $M_{n}(%
\mathcal{A})$.

It follows from Formula (11) that :

\begin{equation}
U(n;\mathcal{A})=\{a\in Inv(M_{n}(\mathcal{A}))/a^{*}=a^{-1}\}.  \tag{13}
\label{13}
\end{equation}

By (11) and Theorem 6.1, its Lie algebra $\mathcal{U}_{\mathbb{I}_{n}}(n;%
\mathcal{A}))$, that we denote by $\mathcal{U}(n;\mathcal{A}))$ is :

\begin{equation}
\mathcal{U}(n;\mathcal{A}))=\{a\in M_{n}(\mathcal{A})/a^{*}+a=0\}.  \tag{14}
\label{14}
\end{equation}

In the case $n=1$, one usually lets $\mathcal{U}(\mathcal{A})=\mathcal{U}(1;%
\mathcal{A})$ et $U(\mathcal{A})=U(1,\mathcal{A})\,$and one has then :

\begin{equation}
\left\{ 
\begin{array}{l}
\mathcal{U}(\mathcal{A})=\{a\in \mathcal{A}/a^{*}=-a\} \\ 
U(\mathcal{A})=\{a\in \mathcal{A}/a^{*}.a=a.a^{*}=\mathbb{I}\}
\end{array}
\right\} .  \tag{15}  \label{15}
\end{equation}

In the case $n=2$ one obtains :

\begin{equation}
\mathcal{U}(2\ ;\mathcal{A})=\left\{ \left( 
\begin{array}{ll}
a & b \\ 
-b^{*} & d
\end{array}
\right) \in M_{2}(\mathcal{A})/(a,d)\in \mathcal{U}(\mathcal{A})^{2},b\in 
\mathcal{A}\right\}  \tag{16}  \label{16}
\end{equation}

\begin{equation}
U(2\ ;\mathcal{A})=\left\{ \left( 
\begin{array}{ll}
a & b \\ 
c & d
\end{array}
\right) \in M_{2}(\mathcal{A})/\left[ 
\begin{array}{c}
a^{*}.a+c^{*}.c=\mathbb{I} \\ 
b^{*}.b+d^{*}.d=\mathbb{I} \\ 
a^{*}.b+c^{*}.d=0
\end{array}
\right] \right\}  \tag{17}  \label{17}
\end{equation}

b) \textit{The sympletic groups }$\mathcal{S}p(n;\mathcal{A})$. For any
integer $n>0$ the matrix :

\begin{equation*}
\Omega (n)=\left( 
\begin{array}{ll}
0 & -\mathbb{I}_{n} \\ 
\mathbb{I}_{n} & \ \ 0
\end{array}
\right)
\end{equation*}
lies in $\mathcal{J}^{-}(2n;\mathcal{A})$.

We shall denote by $S\mathbf{p}(n;\mathcal{A})$ the Fr\'{e}chet Lie group of
Campbell-Baker-Hausdorff type $U_{\Omega (n)}(2n;\mathcal{A})$. By theorem
6.1, its Lie algebra is $\mathcal{S}p(n;\mathcal{A})=\mathcal{U}_{\Omega
(n)}(2n;\mathcal{A})$. $S\mathbf{p}(n;\mathcal{A})$ will be called the
sympletic group of $M_{2n}(\mathcal{A})=M_{2}(M_{n}(\mathcal{A}))$. Taking
into account Formula (11) an easy computation gives :

\begin{equation}
\mathcal{S}p(n;\mathcal{A})=\left\{ \left( 
\begin{array}{ll}
a & b \\ 
c & -a^{*}
\end{array}
\right) \in M_{2}(M_{n}(\mathcal{A}))/\left[ 
\begin{array}{c}
b^{*}=b \\ 
c^{*}=c
\end{array}
\right] \right\}  \tag{18}  \label{18}
\end{equation}

\begin{equation}
S\mathbf{p}(n;\mathcal{A})=\left\{ \left( 
\begin{array}{ll}
a & b \\ 
c & d
\end{array}
\right) \in M_{2}(M_{n}(\mathcal{A}))/\left[ 
\begin{array}{c}
c^{*}.a=a^{*}.c \\ 
d^{*}.b=b^{*}.d \\ 
d^{*}.a+b^{*}.c=\mathbb{I}_{n}
\end{array}
\right] \right\}  \tag{19}  \label{19}
\end{equation}

c) \textit{The pseudo-unitary groups }$U(p,q;\mathcal{A})$\textit{.}

Let us fix now a pair $(p,q)$ of strictly positive integers : the matrix :

\begin{equation*}
I(p,q)=\left( 
\begin{array}{ll}
\mathbb{I}_{p} & \ \ 0 \\ 
0 & -\mathbb{I}_{q}
\end{array}
\right)
\end{equation*}
lies in $\mathcal{J}^{+}(p+q;\mathcal{A})$. We observe now that for any pair 
$(r,s)$ of strictly positive integers, and denoting by $M_{r,s}(\mathcal{A})$
be the $\mathcal{A}$-module of $r\times s$ matrices over $\mathcal{A}$, so
that $M_{r,r}(\mathcal{A})=M_{r}(\mathcal{A})$, the involution $*$ on $%
\mathcal{A}$ extends into a mapping, still denoted by $*$, from $M_{r,s}(%
\mathcal{A})$ onto $M_{s,r}(\mathcal{A})$ such that :

\begin{equation*}
\left( (a_{i,j})_{1\leq i\leq r,1\leq j\leq s}\right)
^{*}=(a_{j,i}^{*})_{1\leq i\leq r,1\leq j\leq s}
\end{equation*}

The pseudo-unitary group $U(p,q;\mathcal{A})$ is by definition the
Fr\'{e}chet Lie group of Campbell-Baker-Hausdorff type $U_{I(p,q)}(p+q;%
\mathcal{A})$ ; by Theorem 6.1 its Lie algebra is $\mathcal{U}_{I(p,q)}(p+q;%
\mathcal{A})$ that we shall denote by $\mathcal{U}(p+q;\mathcal{A})$. Using
Formula (11) an easy computation gives :

\begin{equation}
\mathcal{U}(p,q;\mathcal{A})=\left\{ \left( 
\begin{array}{ll}
a & b \\ 
b^{*} & d
\end{array}
\right) \in M_{p+q}(\mathcal{A})/\left[ 
\begin{array}{c}
a\in \mathcal{U}(p;\mathcal{A}) \\ 
d\in \mathcal{U}(q;\mathcal{A}) \\ 
b\in M_{p,q}(\mathcal{A})
\end{array}
\right] \right\}  \tag{20}  \label{20}
\end{equation}

\begin{equation}
U(p,q;\mathcal{A})=\left\{ \left( 
\begin{array}{ll}
a & b \\ 
c & d
\end{array}
\right) \in M_{p+q}(\mathcal{A})/\left[ 
\begin{array}{c}
a\in M_{p}(\mathcal{A}) \\ 
b\in M_{p,q}(\mathcal{A}) \\ 
c\in M_{q,p}(\mathcal{A}) \\ 
d\in M_{q}(\mathcal{A}) \\ 
a^{*}.a-c^{*}.c=\mathbb{I}_{p} \\ 
d^{*}.d-b^{*}.b=\mathbb{I}_{q} \\ 
a^{*}.b=c^{*}.d
\end{array}
\right] \right\}  \tag{21}  \label{21}
\end{equation}

\section{Unimodularity in good spectral triples}

\subsection{$\mathcal{K}$-traces}

\begin{definition}
Let $\mathcal{K}=(\mathcal{A},\mathcal{H},D)$ be a good spectral triple.

(1) : By $\mathcal{K}$-trace over $\mathcal{A}$ it will be meant any linear
form $T:\mathcal{A}\rightarrow \mathbb{C}$ on $\mathcal{A}$ such that : 
\begin{eqnarray*}
(\alpha ) &:&T(a.b)=T(b.a),T(a^{*})=\overline{T(a)},T(a^{*}.a)\geq 0\text{
for every }a,b\text{ in }\mathcal{A}\text{, and }T(\mathbb{I})=1; \\
(\beta ) &:&T\text{ is continuous with respect to the }\mathcal{K}\text{%
-topology of }\mathcal{A}.
\end{eqnarray*}

(2) : We shall say that $\mathcal{K}$ is unimodularizable if the set $%
\mathcal{T}(\mathcal{K})$ of $\mathcal{K}$-traces$\,$over $\mathcal{A}$ is
non empty.
\end{definition}

Proposition 7.1 below gives a first class of examples of unimodularizable
good spectral algebras. We recall that given a spectral triple $(\mathcal{A},%
\mathcal{H},D)$, $\left\| .\right\| $ denotes the restriction to $\mathcal{A}
$ of the $C^{*}$-norm of the $C^{*}$-algebra $\mathcal{L}(\mathcal{H})$ of
bounded linear operators on $\mathcal{H}$, and that $\mathcal{A}_{0}$
denotes the $C^{*}$-completion of $\mathcal{A}$ with respect to $\left\|
.\right\| $.

\begin{proposition}
Let $\mathcal{K}=(\mathcal{A},\mathcal{H},D)$ be a good spectral triple such
that the $C^{*}$-algebra $\mathcal{A}_{0}$ has a normalized finite trace $T$%
. The restriction of $T$ to $\mathcal{A}$, still denoted by $T$, is a $%
\mathcal{K}$-trace over $\mathcal{A}$. In particular, if $\mathcal{A}_{0}$
is a factor of type $II_{1}$, the restriction to $\mathcal{A}$ of the unique
normalized faithful finite trace of $\mathcal{A}_{0}$ is a $\mathcal{K}$%
-trace over $\mathcal{A}$.
\end{proposition}

\proof%
%
Let $\mathcal{K}=(\mathcal{A},\mathcal{H},D)$ be a good spectral triple, and
let us assume that the $C^{*}$-algebra $\mathcal{A}_{0}$ has a normalized
finite trace $T$. Then $T$ is continuous, and its restriction to $\mathcal{A}
$, still denoted by $T$, is a linear form on $\mathcal{A}$ which is
continuous with respect to $\left\| .\right\| $ and which fulfils Condition (%
$\alpha $) required in Definition 7.1. We achieve the proof by observing
that, since the topology induced by $\left\| .\right\| $ on $\mathcal{A}$ is
weaker than its $\mathcal{K}$-topology (see proof of Prop. 4.1), $T$ fulfils
also Condition ($\beta $)%
\endproof%
%

\begin{proposition}
Let $\mathcal{K}=(\mathcal{A},\mathcal{H},D)$ be a unimodularizable good
spectral triple, and let $T$ be any $\mathcal{K}$-trace over $\mathcal{A}$.
Then, for any strictly positive integer $n$, the mapping $T_{[n]}:M_{n}(%
\mathcal{A})\rightarrow \mathbb{C}$ defined for any $a=(a_{i,j})_{1\leq
i,j\leq n}$ in $M_{n}(\mathcal{A})$ by : 
\begin{equation*}
T_{[n]}(a)=\frac{1}{n}\overset{i=n}{\underset{i=1}{\sum }}T(a_{i,i})
\end{equation*}
is a $\mathcal{K}_{n}$-trace over $M_{n}(\mathcal{A})$, and then $\mathcal{K}%
_{n}=(M_{n}(\mathcal{A}),\mathcal{H}\otimes \mathbb{C}^{n},D\otimes I_{n})$
is unimodularizable.
\end{proposition}

\proof%
%
It is a direct consequence of Assertion (ii) of Theorem 5.4%
\endproof%
%

c) Given a family $\mathcal{F}=\{\mathcal{K}_{i}=(\mathcal{A}_{i},\mathcal{H}%
_{i},D_{i}),i=1,2,...,n\}$ of $n$ unimodularizable good spectral triples, $n$
being any strictly greater than 1, we shall denote by :

-$\mathcal{A}$ the direct sum of unital $*$-algebras : $\mathcal{A}=\mathcal{%
A}_{1}\oplus \mathcal{A}_{2}\oplus ...\oplus \mathcal{A}_{n}$, so that any
element $a$ in $\mathcal{A}$ is of the form $a=a_{1}\oplus a_{2}\oplus
...\oplus a_{n}$, with $a_{i}$ belonging to $\mathcal{A}_{i}$, $i=1,2,...,n$
;

-$\mathbb{I}=\mathbb{I}_{1}\oplus \mathbb{I}_{2}\oplus ...\oplus \mathbb{I}%
_{n}$ the unit element of $\mathcal{A}$, $\mathbb{I}_{i}$ being the unit
elemnt of $\mathcal{A}_{i}$, $i=1,2,...,n$ ;

-$\mathcal{H\,}$the direct sum of Hilbert spaces : $\mathcal{H}=\mathcal{H}%
_{1}\oplus \mathcal{H}_{2}\oplus ...\oplus \mathcal{H}_{n}$ ;

-$D$ the direct sum of operators : $D_{1}\oplus D_{2}\oplus ...\oplus D_{n}$
(acting naturally on $\mathcal{H}$).

From the fact that all the $\mathcal{K}_{i}=(\mathcal{A}_{i},\mathcal{H}%
_{i},D_{i}),i=1,2,...,n$, are good spectral triples, one easily deduces that
their sum $\mathcal{K}=(\mathcal{A},\mathcal{H},D)$ is a good spectral
triple.

\begin{proposition}
Let $\{\mathcal{K}_{i}=(\mathcal{A}_{i},\mathcal{H}_{i},D_{i}),i=1,2,...,n\}$
be a $\mathcal{F}$- family of $n$ good spectral triples, where $n$ is an
integer strictly greater than 1, and for each index $i=1,2,...,n$, let $%
T_{i} $ be a $\mathcal{K}_{i}$-trace over $\mathcal{A}_{i}$. Let us denote
by $T^{(i)}$, $i=1,2,...,n$, the mapping from $\mathcal{A}$ into $\mathbb{C}$
defined for any element $a=a_{1}\oplus a_{2}\oplus ...\oplus a_{n}$ in $%
\mathcal{A}$ by : 
\begin{equation}
T^{(i)}(a_{1}\oplus a_{2}\oplus ...\oplus a_{n})=T_{i}(a_{i}),i=1,2,...,n. 
\tag{\#}
\end{equation}

Then $\mathcal{K}=(\mathcal{A},\mathcal{H},D)$ is an unimodularizable good
spectral triple, and $\mathcal{T}(\mathcal{K})$ contains the set of elements
of the form $T=u_{1}T_{1}+u_{2}T_{2}+...+u_{n}T_{n}$, $%
(u_{1},u_{2},...,u_{n})$ being any $n$-uplet of non negative real numbers
fulfilling $u_{1}+u_{2}+...+u_{n}=1$.
\end{proposition}

\proof%
%
(i) : One easily deduces from Proposition 7.2 that for any $i=1,2,...,n$,
the mapping $T^{(i)}:\mathcal{A}\rightarrow \mathbb{C}$ defined by (\#) is a 
$\mathcal{K}$-trace over $\mathcal{A}$.

Observing that the system $(T^{(1)},T^{(2)},...,T^{(n)})$ is a linearly
independant system in the vector space $\mathcal{V}(\mathcal{K})$ of linear
forms on $\mathcal{A}$ fulfilling conditions ($\alpha $) and ($\beta $)
given in Definition 6 and which are continuous with respect to the $\mathcal{%
K}$-topology of $\mathcal{A}$, one deduces now that the convex hull of the
set $(T^{(1)},T^{(2)},...,T^{(n)})$ is a subset of the set of $\mathcal{K}$%
-traces over $\mathcal{A}$%
\endproof%
%

\subsection{$\mathcal{K}$-traces and Dixmier traces : the case of finite
direct sums of dense unital $*$-algebras of factors of type $II_{1}$}

a) Let us recall firstly what it is called by A. Connes a $d^{+}$-summable
spectral triple, $d$ being a strictly positive real number (see e.g. [5]).

For any element $L$ in the two-sided ideal $\mathcal{K}(\mathcal{H})$ of
compact operators on some separable Hilbert space $\mathcal{H}$ let us
denote by $(\mu _{n}(L))_{n}$ the sequence of eigenvalues of operator $%
\left| L\right| =(L^{*}L)^{\frac{1}{2}}$ counted with their multiplicities
and arranged in decreasing order. The Dixmier ideal $\pounds ^{1,+}(\mathcal{%
H})$ consists of element $L$ in $\mathcal{K}(\mathcal{H})$ for which the
sequence :

\begin{equation*}
\left( \frac{1}{\log N}\sum_{i=0}^{i=N-1}\mu _{n}(L)\right) _{N\geq 1}
\end{equation*}
is bounded. $\pounds ^{1,+}(\mathcal{H})$ is the domain of the Dixmier trace 
$T_{\omega }$, which allows to sum logarithmic divergences. According to
[5], given a real number $d>0$, a good spectral triple $\mathcal{K}=(%
\mathcal{A},\mathcal{H},D)$ which is such that $\left| D\right| ^{-d}\in
\pounds ^{1,+}(\mathcal{H})$ is said to be a $d^{+}$-summable. It is known
(see e.g. [5]) that for any $d^{+}$-summable spectral triple $\mathcal{K}=(%
\mathcal{A},\mathcal{H},D)$, the associated functional :

\begin{equation*}
\Phi ^{D}:L\longmapsto \Phi ^{D}(L)=T_{\omega }(L.\left| D\right|
^{-d}),L\in \pounds (\mathcal{H})
\end{equation*}
is an hypertrace on $\mathcal{A}$, which means that $\Phi ^{D}(a.L)=\Phi
^{D}(L.a)$, $L\in \pounds (\mathcal{H})$, $\forall a\in \mathcal{A}$, and it
is a trace on $\pounds (\mathcal{H})$ vanishing on the subspace of
Hilbert-Schmidt operators.

We shall denote by $\Xi ^{D}$ the trace on $\mathcal{A}$ obtained by
restriction $\Phi ^{D}$ to $\mathcal{A}$ ; by misuse of language we shall
call $\Xi ^{D}$ : \textit{the Dixmier trace of} $\mathcal{A}$. Of course, if 
$\mathcal{H}$ is a finite-dimensional vector space $\Xi ^{D}$ is the zero
trace, while if $\mathcal{H}$ is an infinite-dimensional vector space, the
unit $\mathbb{I}$ (which is the identity on $\mathcal{H}$) is not
Hilbert-Schmidt, so that $\Xi ^{D}(\mathbb{I})\neq 0$.

\begin{lemma}
Let $\mathcal{K}=(\mathcal{A},\mathcal{H},D)$ be a $d^{+}$-summable good
spectral triple for some strictly positive real number $d$, and let us
assume that $\mathcal{H}$ is an infinite-dimensional Hilbert space and that $%
\mathcal{A}_{0}$ is a factor of type $II_{1}$. Then : 
\begin{equation*}
T=\frac{1}{\Xi ^{D}(\mathbb{I})}\Xi ^{D}
\end{equation*}
is a $\mathcal{K}$-trace over $\mathcal{A}$.
\end{lemma}

\proof%
%
Let us assume that $\mathcal{A}_{0}$ is a factor of type $II_{1}$. Then, the
center of $\mathcal{A}$, like the center of $\mathcal{A}_{0}$ is exactly $%
\mathbb{CI}$ ; taking into account that any trace on $\mathcal{A}$
necessarily vanishes on non central elements, it follows that the Dixmier
trace $\Xi ^{D} $ of $\mathcal{A}$ is a scalar multiple of the restriction
to $\mathcal{A}$ of the unique normalized faithful finite trace of $\mathcal{%
A}_{0}$, and since $\mathcal{H}$ is assumed to be an infinite-dimensional
space, one deduces that $\Xi ^{D}(\mathbb{I})\neq 0$ and then, by
Proposition 7.1, that :

\begin{equation*}
T=\frac{1}{\Xi ^{D}(\mathbb{I})}\Xi ^{D}
\end{equation*}
is a $\mathcal{K}$-trace over $\mathcal{A}$%
\endproof%
%

\subsection{Fr\'{e}chet unimodular groups in unimodularizable good spectral
triples}

\begin{theorem}
Let $\mathcal{K}=(\mathcal{A},\mathcal{H},D)$ be an unimodularizable good
spectral triple ; we endow $\mathcal{A}$ with its $\mathcal{K}$-topology.

(i) : For any $\mathcal{K}$-trace $T$ over $\mathcal{A}$, the space $%
\mathcal{S}_{T}(\mathcal{A})=\ker (T)$ is a closed Lie subalgebra of $%
\mathcal{A}$ containing the derived algebra $[\mathcal{A},\mathcal{A}]$, and
one has the direct topological sum : 
\begin{equation*}
\mathcal{A}=\mathcal{S}_{T}(\mathcal{A})\oplus \mathbb{CI}
\end{equation*}

(ii) : Let $T,T^{\prime }$ be two elements in $\mathcal{T}(\mathcal{K})$ ;
then : $\mathcal{S}_{T}(\mathcal{A})=\mathcal{S}_{T^{\prime }}(\mathcal{A}%
)\Leftrightarrow T=T^{\prime }$.

(iii) : Let $T$ be any $\mathcal{K}$-trace over $\mathcal{A}$. For any
integer $n>0$, there exists an unique connected Fr\'{e}chet Lie subgroup $%
S_{T}(n$ ; $\mathcal{A})$ of Campbell-Baker-Hausdorff type of $Inv(M_{n}(%
\mathcal{A}))$, the Lie algebra of which is $\mathcal{S}_{T}(n;\mathcal{A}%
)=\ker (T_{[n]})$, where $T_{[n]}$ denotes the $\mathcal{K}_{n}$-trace over $%
M_{n}(\mathcal{A})$ defined in Proposition 7.2.

$S_{T}(n$ ; $\mathcal{A})$ will be called the $(\mathcal{K}_{n}$ ; $T)$%
-unimodular group.
\end{theorem}

\proof%
%
(i) : Let $T$ be a $\mathcal{K}$-trace over $\mathcal{A}$. The closeness of $%
\mathcal{S}_{T}(n;\mathcal{A})=\ker (T)$ follows from the continuity of $T$
on $\mathcal{A}$ with respect to the $\mathcal{K}$-topology. Moreover, given
any pair $(a,b)$ of elements on $\mathcal{A}$, since one has $T(b.a)=T(a.b)$%
, one deduces that $T([a,b])=0$ and then, that $\mathcal{S}_{T}(\mathcal{A})$
is a Lie subalgebra of $\mathcal{A}$ containing the derived algebra $[%
\mathcal{A},\mathcal{A}]$.

Endly, observing that for any element $a$ in $\mathcal{A}$ one has :

\begin{equation}
a=a-T(a)\mathbb{I}+T(a)\mathbb{I}  \tag{Y}
\end{equation}
with $T(a-T(a)\mathbb{I})=T(a)-T(a)=0$, one deduces that one has the
topological direct sum :

\begin{equation*}
\mathcal{A}=\mathcal{S}_{T}(\mathcal{A})\oplus \mathbb{CI}.
\end{equation*}

(ii) : Let $T,T^{\prime }$ be two $\mathcal{K}$-traces over $\mathcal{A}$ ;
of course, if $T=T^{\prime }$ one has $\mathcal{S}_{T}(\mathcal{A})=\mathcal{%
S}_{T^{\prime }}(\mathcal{A})$. Conversely, if $\mathcal{S}_{T}(\mathcal{A})=%
\mathcal{S}_{T^{\prime }}(\mathcal{A})$ from $(Y)$ one deduces that :

\begin{equation*}
0=T^{\prime }(a-T(a)\mathbb{I})=T^{\prime }(a)-T(a)T^{\prime }(\mathbb{I}%
)=T^{\prime }(a)-T(a),
\end{equation*}
which proves that $T^{\prime }=T$.\\

(iii) : By Proposition 7.2, $\mathcal{K}_{n}=(M_{n}(\mathcal{A}),\mathcal{H}%
\otimes \mathbb{C}^{n},D\otimes \mathbb{I}_{n})$ is an unimodularizable good
spectral triple (see Theorem 5.4), and $T_{[n]}$ is a $\mathcal{K}_{n}$%
-trace over $M_{n}(\mathcal{A})$ endowed with the $\mathcal{K}_{n}$-topology.

It follows then from (i) that :

\begin{equation*}
\mathcal{S}_{T}(n;\mathcal{A})=\ker (T_{[n]})
\end{equation*}
is a closed Lie algebra of codimension 1 of the Fr\'{e}chet Lie algebra $%
M_{n}(\mathcal{A})$.

By Theorem 5.5, there exists an unique connected Lie subgroup $S_{T}(n$ ; $%
\mathcal{A})$ of Campbell-Baker-Hausdorff type of the Fr\'{e}chet Lie group $%
Inv(M_{n}(\mathcal{A}))$ the Lie algebra of which is $\mathcal{S}_{T}(n;%
\mathcal{A})=\{a\in M_{n}(\mathcal{A})/T_{[n]}(a)=0\}$. Of course, $S_{T}(n$
; $\mathcal{A})$ is the closed connected subgroup of $Inv(M_{n}(\mathcal{A}%
)) $ generated by the elements of the form $Exp_{(n)}v$ (see Theorem 5.5),
with $v$ running in the Lie algebra $\mathcal{S}_{T}(n;\mathcal{A})$%
\endproof%
%

\textbf{Remark 7.1. }(1) : By Assertion (ii) of Theorem 7.5., two different $%
\mathcal{K}$-traces $T$ and $T^{\prime }$ over $\mathcal{A}$ give rise, for
any integer $n>0$, to two different unimodular groups $S_{T}(n$ ; $\mathcal{A%
})$ and $S_{T^{\prime }}(n$ ; $\mathcal{A})$.\\

(2) : In [5] the notion of unimodular transformation, in the context of a
Von Neumann algebra of type $II_{1}$ has been already considered, using the
natural normalized trace, which is really a continuous trace for the $%
C^{\ast }$-algebra structure ; our approach is the same.

Neverthless, if we want that the unimodular transformations of non Banach
Fr\'{e}chet $*$-algebra $\mathcal{A}$ with respect to some continuous trace $%
T$ (if it exists) constitutes a Lie group like in the Banach case, it is
necessary that the closed Lie algebra $Ker(T)$ is an integrable Lie
subalgebra of $\mathcal{A}$ ; unfortunately, since the works of H. Omori, it
is known that a closed Lie subalgebra of the Lie algebra of a Fr\'{e}chet
Lie group is not necessarily integrable. The fact that in a good spectral
triple $\mathcal{K}=(\mathcal{A},\mathcal{H},D)$ the group $Inv(\mathcal{A})$
is of Campbell-Baker-Hausdorff type is then very important for the
integrability of $Ker(T)$.\\

(3) : Lemma 7.4 shows that some $d^{+}$-summable good spectral triple are
unimodularizable. It should be interesting to charaterize, in the general
case, the $d^{+}$-summable good spectral triples which are unimodularizable.\\

Given any good spectral triple $\mathcal{K}=(\mathcal{A},\mathcal{H},D)$,
any strictly positive integer $n$, and any element $\Omega $ in $\mathcal{J}%
(n;\mathcal{A})=\{\Omega \in Inv(M_{n}(\mathcal{A}))/\Omega ^{-1}=\Omega
^{*}=\varepsilon \Omega ,\varepsilon \in \{-1,1\}\}$, in paragraph 5.1 we
have seen that $U_{\Omega }(n;\mathcal{A})=\{a\in Inv(M_{n}(\mathcal{A}%
))/a^{*}.\Omega .a=\Omega \}$ is a real closed Fr\'{e}chet Lie subgroup of
Campbell-Baker-Hausdorff type of $Inv(M_{n}(\mathcal{A}))$, the Lie algebra
of which is the real Lie algebra (Theorem 6.1) :

\begin{equation*}
\mathcal{U}_{\Omega }(n;\mathcal{A})=\{a\in M_{n}(\mathcal{A})/a^{*}.\Omega
+\Omega .a=0\}.
\end{equation*}

We are now able to precise what are the unimodular subgroups of these
groups. More precisely :

\begin{theorem}
Let $\mathcal{K}=(\mathcal{A},\mathcal{H},D)$ be an unimodularizable good
spectral triple, and let $T$ be any $\mathcal{K}$-trace over $\mathcal{A}$.
For any integer $n>0$ and any element $\Omega $ in $\mathcal{J}(n;\mathcal{A}%
)$ there exists an unique connected Fr\'{e}chet Lie subgroup $S_{T}U_{\Omega
}(n$ ; $\mathcal{A})$ of Campbell-Baker-Hausdorff type of $U_{\Omega }(n$ ; $%
\mathcal{A})$ the Lie algebra of which is : 
\begin{equation*}
\mathcal{S}_{T}\mathcal{U}_{\Omega }(n;\mathcal{A})=\{a\in \mathcal{S}_{T}(n;%
\mathcal{A})/a^{*}.\Omega +\Omega .a=0\}.
\end{equation*}
\end{theorem}

\proof%
%
Let $\mathcal{S}_{T}\mathcal{U}_{\Omega }(n$ ; $\mathcal{A})$ be the set $%
\{x\in \mathcal{U}_{\Omega }(n;\mathcal{A})/T_{[n]}(x)=0\}$, $T_{[n]}$ being
the $\mathcal{K}_{n}$-trace over $M_{n}(\mathcal{A})$ defined in Proposition
7.2. One has then :

\begin{equation*}
\mathcal{S}_{T}\mathcal{U}_{\Omega }(n;\mathcal{A})=\mathcal{U}_{\Omega }(n;%
\mathcal{A})\bigcap \mathcal{S}_{T}(n;\mathcal{A}).
\end{equation*}

It follows then, by Theorem 7.5 and the continuity of $T_{[n]}$, that $%
\mathcal{S}_{T}\mathcal{U}_{\Omega }(n;\mathcal{A})$ is a closed real Lie
subalgebra of the Lie algebra $\mathcal{S}_{T}(n;\mathcal{A})$ of the $(%
\mathcal{K}_{n},T)$-unimodular group, and then of the Lie algebra $M_{n}(%
\mathcal{A})$ of the Fr\'{e}chet Lie group of Campbell-Baker-Hausdorff type $%
Inv(M_{n}(\mathcal{A}))$. By Theorem 5.5(ii) it follows that there exists an
unique connected Lie subgroup $S_{T}U_{\Omega }(n$ ; $\mathcal{A})$ of
Campbell-Baker-Hausdorff type of $Inv(M_{n}(\mathcal{A}))$ whose Lie algbera
is the real Lie Fr\'{e}chet algebra $\mathcal{S}_{T}\mathcal{U}_{\Omega }(n$
; $\mathcal{A})$. Since $\mathcal{S}_{T}\mathcal{U}_{\Omega }(n$ ; $\mathcal{%
A})$ is a closed Lie algebra of $\mathcal{U}_{\Omega }(n$ ; $\mathcal{A})$
one easily deduces that $S_{T}U_{\Omega }(n$ ; $\mathcal{A})$ is a real
connected closed Lie subgroup of $U_{\Omega }(n$ ; $\mathcal{A})$ and of the 
$(\mathcal{K}_{n},T)$-unimodular group $S_{T}(n;\mathcal{A})$%
\endproof%
%

\subsection{Examples of unimodular unitary groups}

a) Let be a family of $n$ unimodularizable good spectral triples $\mathcal{F}%
=\{\mathcal{K}_{i}=(\mathcal{A}_{i},\mathcal{H}_{i},D_{i}),i=1,2,...,n\}$, $%
n $ being any strictly greater than 1, and for each $i=1,2,...,n$, let $%
T_{i} $ be a $\mathcal{K}_{i}$-trace over $\mathcal{A}_{i}$. For each index $%
k=1,2,...,n$, we shall denote by $U(\mathcal{A}_{k})_{0}$ the connected
component of the unit of the unitary group $U(\mathcal{A}_{k})$ of $\mathcal{%
A}_{k}$. By Proposition 7.3, one has that :

\begin{equation*}
\mathcal{K}=(\mathcal{A}_{1}\oplus \mathcal{A}_{2}\oplus ...\oplus \mathcal{A%
}_{n},\mathcal{H}=\mathcal{H}_{1}\oplus \mathcal{H}_{2}\oplus ...\oplus 
\mathcal{H}_{n},D_{1}\oplus D_{2}\oplus ...\oplus D_{n})
\end{equation*}
is an unimodularizable good spectral triple.

Moreover, denoting, for each $i=1,2,...,n$, by $T^{(i)}$ the $\mathcal{K}$%
-trace over $\mathcal{A}$ obtained by taking, for any element $a=a_{1}\oplus
a_{2}\oplus ...\oplus a_{n}$ in $\mathcal{A}:$

\begin{equation*}
T^{(i)}(a_{1}\oplus a_{2}\oplus ...\oplus a_{n})=T_{i}(a_{i})\text{, }%
i=1,2,...,n,
\end{equation*}
and one knows, by Prop. 7.3, that given any $n$-uplet $%
u=(u_{1},u_{2},...,u_{n})$ of positive real numbers such that $%
u_{1}+u_{2}+..+u_{n}=1$, then, $%
T^{[u]}=u_{1}T^{(1)}+u_{2}T^{(2)}+...+u_{n}T^{(n)}$ is $\mathcal{K}$-trace
over $\mathcal{A}$.

One easily deduces, for each $i=1,2,...,n$, that :

\begin{equation*}
\mathcal{S}_{T^{(i)}}\mathcal{U}_{\Omega }(\mathcal{A})=\{a=a_{1}\oplus
a_{2}\oplus ...\oplus a_{n}\in \mathcal{A}/a_{k}+a_{k}^{*}=0\quad \forall
k=1,2,...,n\text{, and }T_{i}(a_{i})=0\}
\end{equation*}
is the real Lie Fr\'{e}chet algebra of the unimodular unitary group of $%
\mathcal{A}$ with respect to the $\mathcal{K}$-trace $T^{(i)}:$

\begin{equation*}
S_{T^{(i)}}U_{\Omega }(\mathcal{A})=U(\mathcal{A}_{1})_{0}\oplus ...\oplus U(%
\mathcal{A}_{i-1})_{0}\oplus S_{T_{i}}U_{\Omega }(\mathcal{A}_{i})\oplus U(%
\mathcal{A}_{i+1})_{0}\oplus ...\oplus U(\mathcal{A}_{n})_{0}
\end{equation*}
$S_{T_{i}}U_{\Omega }(\mathcal{A}_{i})$ being the unimodular unitary group
of $\mathcal{A}_{i}$ w.r.t. $T_{i}$, the Lie algebra of which is :

\begin{equation}
\mathcal{S}_{T_{i}}\mathcal{U}_{\Omega }(\mathcal{A}_{i})=\{a_{i}\in 
\mathcal{A}_{i}/a_{i}+a_{i}^{*}=0\text{ and }T_{i}(a_{i})=0\}.  \tag{23}
\label{23}
\end{equation}

More generally, given any element $u$ in the set :

\begin{equation*}
C_{n}=\{u=(u_{1},u_{2},...,u_{n})\in \mathbb{R}^{n}/u_{i}\geq 0\text{, }%
i=1,2,...,n\text{, and }u_{1}+u_{2}+...+u_{n}=1\}
\end{equation*}
the unimodular unitary group $S_{T^{[u]}}U(\mathcal{A})$ is the connected
Fr\'{e}chet Lie subgroup of $U(\mathcal{A})$ with Lie algebra : 
\begin{equation*}
\mathcal{S}_{T^{[u]}}\mathcal{U}(\mathcal{A})=\{a=a_{1}\oplus a_{2}\oplus
...\oplus a_{n}\in \mathcal{A}%
/u_{1}T_{1}(a_{1})+u_{2}T_{2}(a_{2})+..+u_{n}T_{n}(a_{n})=0\}.
\end{equation*}

If $u$ and $v$ are two different elements in $C_{n}$, then taking into
account Remark 7.1, one has that :

\begin{equation*}
\mathcal{S}_{T^{[u]}}\mathcal{U}(\mathcal{A})\neq \mathcal{S}_{T^{[v]}}%
\mathcal{U}(\mathcal{A}).
\end{equation*}

\newpage

\textbf{Acknowledgments}{\large . }The authors express their gratitude to H.
Omori for his suggestions made during his participation at the Colloquium
''Analysis on infinite-dimensional Lie groups and algebras'' which held in
Marseille in September 1997, to D. Testard whose many constructive
criticisms were used in the part of this paper related to the unimodularity,
and to D. Kastler who gave us the opportunity to give a talk on our work in
the workshop ''Quantum groups'' of Palermo in December 1997.


\begin{thebibliography}{10}
\bibitem[1]{1}  B. Blackadar and J. Cuntz : Differential Banach algebras
norms and smooth subalgebras of $C^{*}$-algebras, J. Operator Theory 26,
(1991), 255-282.

\bibitem[2]{2}  J. Bochnak and S. Siciak : Analytic functions in topological
vector spaces, Studia Math. 39, Math. 101, (1990), 261-333.

\bibitem[3]{3}  J.-B. Bost : Principe d'Oka, $K$-th\'{e}orie et syst\`{e}mes
dynamiques non commutatifs, Invent. Math. 101, (1990), 261-333.

\bibitem[4]{4}  N. Bourbaki : \textit{Groupes et alg\`{e}bres de Lie}, Chap.
2 et 3, Hermann, Paris, 1972.

\bibitem[5]{5}  A. Connes : \textit{Non Commutative Geometry}, Academic
Press, New York, 1994.

\bibitem[6]{6}  J. Leslie : Some integrable subalgebras of the Lie algebras
of infinite-dimensional Lie groups, Trans. of the A.M.S., Vol 333, (1992),
423-443.

\bibitem[7]{7}  L. Loomis : \textit{An introduction to abstract harmonic
analysis}, Van Nostrand, New York, 1853.

\bibitem[8]{8}  J. Marion : Energy representations of infinite dimensional
gauge groups in non commutative geometry, International Journal of
Mathematics, Vol.5, N${{}^{\circ }}3$ (1994), 329-348.

\bibitem[9]{9}  J. Marion : Gauge frameworks over locally $m$-convex
Fr\'{e}chet unital involutive algebras and gauge representations of their
unitary groups, to appear in ''Methods of Functional Analysis and
Topology'', Kiev, 1998.

\bibitem[10]{10}  J. Marion and T. Robart : Regular Fr\'{e}chet-Lie group
structures of Campbell-Baker-Hausdorff type and Lie's second fundamental
theorem in inverse limits of unital involutive Banach algebras, Georgian
Mathematical Journal, Vol. 2, N${{}^{\circ }}$4 (1995), 425-444.

\bibitem[11]{11}  J. Milnor : Remarks on infinite dimensional Lie groups, in
Proceeding of Summer School on quantum gravity, Les Houches, Ed. B. S.
Dewitt and R. Stora, Elsevier Sc. Pub., (1984), 1008-1057.

\bibitem[12]{12}  H. Omori : \textit{Infinite dimensional Lie groups},
Trans. of Math. Monographs, A.M.S., Providence, 1997.

\bibitem[13]{13}  T. Robart : Groupes et alg\`{e}bres de Lie de dimension
infiinie : probl\`{e}mes d'int\'{e}grabilit\'{e} et de r\'{e}gularit\'{e},
Ph.D. dissertion, Faculty of Sciences, University of Aix-Marseille II, 1994.

\bibitem[14]{14}  R. Swan : Topological examples of projectives modules,
Trans. Am. Math. Soc. 230, (1977), 201-234.
\end{thebibliography}
\end{document}